\documentclass[12pt]{article}
\usepackage{amsmath,epsfig,euscript}

\def\calAslash{\rlap{\hspace{0.08cm}/}{{\EuScript A}}}
\def\calDslash{\rlap{\hspace{0.1cm}/}{{\EuScript D}}}
\def\delslash{\rlap{\hspace{0.02cm}/}{\partial}}
\def\Dslash{\rlap{\hspace{0.07cm}/}{D}}
\def\nslash{\rlap{\hspace{0.02cm}/}{n}}
\def\nbslash{\rlap{\hspace{0.02cm}/}{\bar n}}
\def\vslash{\rlap{\hspace{0.02cm}/}{v}}

\def\A{{\EuScript A}}

\def\X{{\EuScript X}}

\begin{document}

\begin{titlepage}

\begin{flushright}
{\tt arXiv:0903.3377}\\[0.2cm]
\end{flushright}

\vspace{0.7cm}
\begin{center}
\Large\bf\boldmath
Subleading Jet Functions in Inclusive B Decays
\unboldmath
\end{center}

\vspace{0.8cm}
\begin{center}
{\sc  Gil Paz}\\
\vspace{0.4cm}
{School of Natural Sciences, Institute for Advanced Study\\
Princeton, NJ 08540, U.S.A.}
\end{center}
\vspace{1.0cm}
\begin{abstract}
  \vspace{0.2cm} 
  \noindent 

  The contribution of subleading jet functions to inclusive decay
  distributions of $B$ mesons are derived from a systematic two-step
  matching of QCD current correlators onto soft collinear and heavy
  quark effective theory. Focusing on the tree level matching of QCD
  onto soft collinear effective theory, the subleading jet functions
  are defined to all orders in $\alpha_s(\mu_i)$ (with $\mu_i^2\sim
  m_b\Lambda_{\rm QCD}$) and are calculated explicitly at first order
  in $\alpha_s(\mu_i)$. We present explicit expressions for the decay
  rates of $\bar B \to X_u\, l\,\bar\nu$ and the $Q_{7\gamma}-Q_{7\gamma}$
  contribution to $\bar B \to X_s\,\gamma$, where the subleading jet
  functions are multiplied by a tree level hard function and appear in
  a convolution with the leading order shape function. Together with
  the recent two loop calculation of the leading order hard function
  for $\bar B \to X_u\, l\,\bar\nu$, this paper will allow for a more
  precise description of inclusive $B$ decays in the end point region.

\end{abstract}
\vfil

\end{titlepage}

\section{Introduction}
Charmless inclusive $B$ decays play an important role in our
understanding of the standard model and its possible extensions. The
inclusive semileptonic $B$ decay currently allows for the most
accurate determination of $|V_{ub}|$, one of the fundamental
parameters of the standard model, while the inclusive $\bar B \to X_s\,
\gamma$ rate is used extensively in constraining models of new
physics.

Since $\Lambda_{\rm QCD}$ is much smaller than the $b$-quark mass
($m_b$), one would expect that various physical observables for
inclusive $B$ decays can be expressed in terms of a local operator
product expansion (OPE), where the various operators are suppressed by
an increasing power of $m_b$. This is the case for the partial and
total rate of $\bar B \to X_c\, l\,\bar\nu$, where schematically we have
\begin{equation}
\label{eq:b2c}
d\Gamma\sim c_{\,0}\,O_{0}+\sum_{i=2}\,\sum_j \frac{1}{m_b^i}\,c_{i}^{\,j}\, O_{i}{^j}.
\end{equation}
The current state of the art is that $c_0$ is known at ${\cal
  O}(\alpha_s^2)$ \cite{Melnikov:2008qs,Pak:2008qt}, while $c_3^{\,j}$
\cite{Gremm:1996df} and $c_4^{\,j}$ \cite{Dassinger:2006md} are known
at ${\cal O}(\alpha_s^0)$. For $c_2^{\, j}$, the coefficient of the
``kinetic energy" operator $c_2^1$, is known at ${\cal O}(\alpha_s)$
\cite{Becher:2007tk}, while $c_2^2$, the coefficient of the
``chromomagnetic" operator, is known only at ${\cal O}(\alpha_s^0)$
\cite{Blok:1993va,Manohar:1993qn}.

For the charmless inclusive $B$ decays, $\bar B \to X_s\, \gamma$ and $\bar B \to
X_u\, l\,\bar\nu$, the situation is more complicated.  Experimental
cuts force the hadronic jet $X$ to have large energy $E_X\sim m_b$,
but only moderate invariant mass $P_X^2\sim m_b\Lambda_{\rm QCD}$.
Consequently there are three energy scales in the problem: a hard
scale ($\mu_h\sim m_b$), a hard-collinear scale ($\mu_i\sim
\sqrt{m_b\Lambda_{\rm QCD}}$) and a soft scale ($\mu_s\sim
\Lambda_{\rm QCD}$). For this kinematical region, often called the
``end point region" or ``the shape function region", the partial rates
can be expressed in terms of a non-local OPE. Thus, for $\bar B \to X_u\,
l\,\bar\nu$ and the $Q_{7\gamma}-Q_{7\gamma}$ contribution to $\bar B \to
X_s\,\gamma$ we have  an expansion which is schematically
\cite{Paz:2006eq}:
\begin{eqnarray}
\label{eq:nlope} d\Gamma_u&\sim& H_u\cdot J\otimes
S+\sum_{i=1}\,\sum_{j,\,k,\,l}
\frac{1}{m_b^i}\,h_i^{j}\cdot j_i^{k}\otimes s_i^{l}\nonumber\\
d\Gamma_s&\sim& H_s\cdot J\otimes S+\sum_{i=1}\,\sum_{j,\,k,\,l}
\frac{1}{m_b^i}\,h_i^{j}\cdot j_i^{k}\otimes s_i^{l},
\end{eqnarray}
where the hard functions ($H$, $h_i$) and the jet functions ($J$,
$j_i$) are calculable in perturbation theory, while the shape
functions ($S$, $s_i$) are non-local light-cone operators which are
non perturbative objects\footnote{For other contributions to
  $d\Gamma_s$, such as $Q_{7\gamma}-Q_{8g}$, one finds that more
  complicated factorization theorems hold
  \cite{Lee:2006wn,Neubert:2008cp,Lee:2008}. We will not discuss these
  contributions in this paper.}. Factorization theorems such as (\ref{eq:nlope}) are most
conveniently proven using the Soft Collinear Effective Theory (SCET)
\cite{Bauer:2000yr,Bauer:2001yt,Beneke:2002ph}.

The current state of the art is as follows. $H_u$ was recently
calculated at ${\cal O}(\alpha_s^2)$ \cite{Bonciani:2008wf,
  Asatrian:2008uk, Beneke:2008ei, Bell:2008ws} and $H_s$
\cite{Neubert:2004dd} is known\footnote{The ${\cal O}(\alpha_s^2)$
  expression can be extracted from known results in the literature
  \cite{Melnikov:2005bx,Blokland:2005uk,Asatrian:2006ph, Asatrian:2006sm}, as was done
  in \cite{Becher:2006pu} for a ``normalized" $H_s$ at the scale
  $\mu_h=m_b$.} at ${\cal O}(\alpha_s)$. The leading order jet
function $J$ is known at ${\cal O}(\alpha_s^2)$ \cite{Becher:2006qw}.
Of the $1/m_b$ corrections, only the terms of the form $h_1^{0}\cdot
j_1^{0}\otimes s_1^{1}$ are explicitly known at ${\cal O}(\alpha_s^0)$
\cite{Lee:2004ja,Bosch:2004cb,Beneke:2004in}. Therefore the
factorization formula was proven only for the leading order term and
one of the $1/m_b$ suppressed terms.

The knowledge of the leading order hard and jet functions at ${\cal
  O}(\alpha_s)$ and the $h_1^{0}\cdot j_1^{0}\otimes s_1^{1}$ terms at
${\cal O}(\alpha_s^0)$ (as well as known, but not properly factorized,
$\alpha_s/m_b^i$ and $1/m_b^2$ corrections) was the basis of the
precision determination of $|V_{ub}|$ in
\cite{Lange:2005yw,Lange:2005qn,Lange:2005xz}.  In order to improve
the accuracy even further, one would like to know as much as possible
about the properly factorized $\alpha_s^2$, $\alpha_s/m_b$, and the
$1/m_b^2$ corrections, in decreasing order of importance.  What would
the calculation of these corrections entail?

In order to find the least important term, namely the $1/m_b^2$
corrections, the heavy-to-light SCET currents need to be matched at
tree level to fourth order in the SCET expansion parameter
$\sqrt{\Lambda_{\rm QCD}/m_b}$. The most important term, namely the
$\alpha_s^2$ corrections, should appear shortly now that the last
ingredient, $H_u$, was calculated at ${\cal O}(\alpha_s^2)$
\cite{Bonciani:2008wf, Asatrian:2008uk, Beneke:2008ei,Bell:2008ws}.
The second most important correction is of order $\alpha_s/m_b$. From
(\ref{eq:nlope}) one would naively expect to find three terms which
scale like $1/m_b$ and need to be calculated to ${\cal O}(\alpha_s)$:
$h^{1}\cdot j^{0}\otimes s^{0}, h^{0}\cdot j^{1}\otimes s^{0}$, and
$h^{0}\cdot j^{0}\otimes s^{1}$ (the subscript $1$ is implicit). Less
formally, we expect to find subleading hard, jet and shape functions,
respectively. Let us discuss each of these terms.

Since the hard functions are products of Wilson coefficients extracted
in the matching of QCD onto SCET, they depend only on kinematical
quantities which scale like $m_b$, i.e. $m_b$ and $E_X+|\vec{P}_X|$.
As such they always scale as ${\cal O}(1)$ in the $1/m_b$ expansion,
so it is clear that terms of the form $h^{1}\cdot j^{0}\otimes s^{0}$
cannot appear at {\it any} order in $\alpha_s$. In other words,
subleading hard functions can appear only when they are multiplied by
subleading jet or shape functions.

The ``subleading shape functions" (SSF), i.e. terms of the form
$h^{0}\cdot j^{0}\otimes s^{1}$ which arise already at ${\cal
  O}(\alpha_s^0)$, can be calculated, in principle, at ${\cal
  O}(\alpha_s)$, i.e.  both $h^{0}$ and $j^{0}$ need to be calculated
at ${\cal O}(\alpha_s)$. For the former a one loop matching of the
SCET current to second order is needed, while the calculation of the
latter was outlined (but not explicitly done!) in \cite{Lee:2004ja} and
\cite{Beneke:2004in}.

The focus of this paper is to prove that terms of the form $h^{0}\cdot
j^{1}\otimes s^{0}$, i.e. ``subleading jet functions" (SJF), indeed
exist in the factorization formula for inclusive $B$ decays. We will
first prove the existence of such terms by showing that the partonic
${\cal O}(\alpha_s)$ terms in the hadronic tensor which are $1/m_b$
suppressed in the end point region arise from two momentum regions:
soft and hard-collinear. We will then show how the soft region is
accounted for by the parton level one loop diagrams of the
\emph{known} ${\cal O}(\alpha_s^0)$ $h^0\cdot j^0\otimes s^1$ term,
and reproduce the hard-collinear region via time order products (TOPs)
of subleading SCET currents. After establishing the need for the
$h^{0}\cdot j^{1}\otimes s^{0}$ term, we will calculate the subleading
jet functions via the usual two step matching.  In the first step the
QCD currents and Lagrangian are matched onto SCET at tree level and to
second order in $\sqrt{\Lambda_{\rm QCD}/m_b}$.  In the second step
the SCET current correlator is matched onto Heavy Quark Effective
Theory (HQET) \cite{Neubert:1993mb} and the subleading jet functions
are extracted.  Since we are interested in $\alpha_s/m_b$ suppressed
terms, and the subleading jet functions start at ${\cal O}(\alpha_s)$,
it is sufficient to consider only the case of tree level matching of
QCD onto SCET, for which we can use known results from the literature.
For this case we will define the subleading jet functions to all
orders in $\alpha_s(\mu_i)$, and calculate them explicitly at first
order in $\alpha_s(\mu_i)$.

The subleading jet functions' contribution is, in some sense, the most
important term at order $\alpha_s/m_b$. When integrating over larger
portions of phase space, the one loop subleading jet functions'
contribution is no longer power suppressed.  The other $\alpha_s/m_b$
term in the factorization formula, namely the ${\cal O}(\alpha_s)$
$h^0\cdot j^0\otimes s^1$ contribution, although formally
$\alpha_s/m_b$ suppressed in the end point region, is expected to
become even more power suppressed and thus is less important outside
of the end point region. In other words, the terms which we will
calculate are kinematically and not hadronically suppressed, and as
such are important outside of the end point region. Incidentally,
experiments are starting to probe the kinematic area outside of the
end point region. Finally, since the subleading jet functions appear
in convolution with the leading order shape function, their inclusion
does not introduce new hadronic uncertainties.

Another motivation for studying subleading jet functions is that they
arise also outside of the context of flavor physics. Since these
functions encode the interaction of hard-collinear quarks and gluons,
the same functions are expected to appear in the $x\to 1$ region of
deep inelastic scattering \cite{Akhoury:1998gs, Akhoury:2003fw,
  Chay:2005rz, Becher:2006mr}.

The rest of the paper is organized as follows. After a short review of
known results in section \ref{sec:review}, we calculate in section
\ref{sec:regions} the partonic ${\cal O}(\alpha_s)$ terms in the
hadronic tensor which are $1/m_b$ suppressed in the end point region.
We then show that they arise from two momentum regions: soft and
hard-collinear.  In section \ref{sec:EFT} we explain how the soft
region is accounted for by the parton level one loop diagrams of the
\emph{known} $h^0\cdot j^0\otimes s^1$ terms, and reproduce the
hard-collinear region via time order products of subleading SCET
currents. In the main section of the paper, section \ref{sec:subjet},
we define the subleading jet functions, for the case of a tree level
hard function, to all orders in $\alpha_s(\mu_i)$ and calculate their
one loop expressions. After a short discussion of their
renormalization, we present properly factorized expressions for the
decay rates of $\bar B \to X_u\, l\,\bar\nu$ and the
$Q_{7\gamma}-Q_{7\gamma}$ contribution to $\bar B \to X_s\,\gamma$. In
section \ref{sec:conclusions} we present our conclusions. The
appendices contain proofs for some of the statements made in section
\ref{sec:subjet}. A reader who is mostly interested in the
phenomenological results, can skip section \ref{sec:regions} and
\ref{sec:EFT} and proceed directly to section \ref{sec:subjet}.

\section{Review}
\label{sec:review} In order to make the paper self-contained, we
review in this section some known results about inclusive $B$ decays,
as well as some basic ingredients of SCET. For a more detailed
account~see~\cite{Paz:2006zz}.

\paragraph{Kinematical Variables}

The kinematics of inclusive $B$ decays is such that in its rest frame,
the $B$ meson decays into a hadronic jet carrying momentum $P_X$ and a
non-hadronic jet (a lepton pair for $\bar B \to X_u\, l\,\bar\nu$ and a
photon for $\bar B \to X_s\,\gamma$) carrying momentum $q$. Denoting by
$M_B$ the mass of the $B$ meson and by $v$ its four-velocity, we have
therefore $M_B v=P_X+q$.  Taking the four velocity of the $B$ meson to
be $v=(1,0,0,0)$ and $\vec{q}$ to point in the negative $z$ direction,
we define two light-like vectors $n^\mu=(1,0,0,1)$,
$\bar{n}^\mu=(1,0,0,-1)$, such that $n+\bar n=2 v$, $n\cdot \bar n=2$,
and $n\cdot v=\bar n\cdot v=1$.  Any four vector $a^\mu$ can be
decomposed as
\begin{equation}
a^\mu=\bar{n}\cdot a\frac{n^\mu}{2}+n\cdot
a\frac{\bar{n}^\mu}{2}+a^\mu_\perp.
\end{equation}
Notice that we have taken $v_\perp=q_\perp=0$. Having fixed the two
light-like vectors, rotational invariance implies that the transverse
indices can only be contracted using
\begin{equation}
g_\perp^{\mu\nu}=g^{\mu\nu}-\frac{n^\mu \bar{n}^\nu+n^\nu \bar{n}^\mu}{2},\quad 
\epsilon_\perp^{\mu\nu}=\frac12\epsilon^{\mu\nu\alpha\beta}\bar{n}_\alpha n_\beta,
\end{equation}
where $\epsilon_{0123}=1$.

Conservation of 4-momentum implies that for the $\bar B \to X_s\,\gamma$
decay mode there is one independent kinematical variable, which we can
take to be the photon energy $E_\gamma$ or $n\cdot P_X=M_B-2E_\gamma$.
The $\bar B \to X_u\, l\,\bar\nu$ decay mode has three independent variables
which we can take to be \cite{Lange:2005yw,Bosch:2004th,Bosch:2004bt}
\begin{equation}
P_+=E_X-|\vec{P}_X|=n\cdot P_X,\quad P_-=E_X+|\vec{P}_X|=\bar n\cdot P_X,\quad P_l=M_B-2E_l.
\end{equation}
These are the ``hadronic" variables. It is also useful to define a
``partonic" set of variables.  Let  $\bar\Lambda=M_B-m_b$,
where $m_b$ is the $b$ quark mass.
Defining $p=m_bv-q=P_X-\bar\Lambda v$, we have $n\cdot p=n\cdot
P-\bar\Lambda,\,\bar n\cdot p=\bar n\cdot P-\bar\Lambda$ as the
corresponding partonic variables to $P_+$ and $P_-$. The hadronic
tensor is naturally expressed in terms of $n\cdot p$ and $\bar n\cdot
p$. Notice also that by construction $p_\perp=0$, and as a result
$p^2=\bar n\cdot p\,\, n\cdot p$.

\paragraph{Hadronic Tensor}

Partial rates for the inclusive decays $\bar B \to X_u\, l\,\bar\nu$, and
the $Q_{7\gamma}-Q_{7\gamma}$ contribution to $\bar B \to X_s\,\gamma$ can
be calculated using the optical theorem. The central object to
consider is the hadronic tensor, which is the discontinuity of a
forward matrix element of a correlator of two currents:
\begin{equation}\label{eq:Wij}
W_{ij}=\frac {1}{\pi}\frac{1}{2M_B}\,{\rm
Im}\,\langle\bar{B}(v)|i\int\,d^4x\, e^{iq\cdot x}\,
T\left\{J^\dagger_i(0)\, J_j(x)\right\}|\bar{B}(v)\rangle.
\end{equation}
where, again, $v$ is the velocity of the B-meson and $q$ is the
momentum of the lepton pair (photon) in the $\bar B \to X_u\, l\,\bar\nu$
($\bar B \to X_s\, \gamma$) decay. The currents can generally be written as
$J^\dagger_i=\bar b\,\Gamma_i\, q$ and $J_j=\bar q\,\Gamma_j\, b$. For
semileptonic decays $\Gamma_i=\gamma^\mu(1-\gamma_5)$ and
$\Gamma_j=\gamma^\nu(1-\gamma_5)$, and for the
$Q_{7\gamma}-Q_{7\gamma}$ contribution to $\bar B \to X_s\, \gamma$,
$\Gamma_i=\frac12(1-\gamma_5)\gamma^\perp_\mu\,\nbslash$,
$\Gamma_j=\frac12{\nbslash}\gamma_\perp^\mu(1-\gamma_5)$. In order to
somewhat simplify the traces in the expression for the hadronic
tensor, we assume that $\Gamma_i$ and $\Gamma_j$ contain the same
number of Dirac's gamma matrices, but otherwise we take $\Gamma_{i,j}$
to be arbitrary Dirac structures.

For radiative decays the hadronic tensor is given in term of one
function $W\equiv W_{ij}$.  The $Q_{7\gamma}-Q_{7\gamma}$ contribution
to photon spectrum can then be written as\footnote{See
  \cite{Lange:2005yw} for the exact definition of the various
  parameters in the this equation.  Notice that $W$ equals
  $-2U(\mu_h,\mu_i)\,{\cal F}_\gamma$ of \cite{Lange:2005yw}. }
\begin{equation}
\label{eq:Wiispec}
\frac{d\Gamma}{dE_\gamma}=-\frac{G_F^2\alpha}{4\pi^4}E_\gamma^3|V_{tb}V^*_{ts}|^2\,\overline{m}_b^2\,|C_{7\gamma}^{\rm eff}|^2\,W(P_+).
\end{equation}
For semileptonic decays we can decompose the hadronic tensor in terms
of five functions, $\tilde{W}_i(P_+,P_-)$,
\begin{eqnarray}
\label{eq:Widef}
W_{ij}=W^{\mu\nu}&=&(n^\mu v^\nu+n^\nu v^\mu-g^{\mu\nu}-i\epsilon^{\mu\nu\alpha\beta} n_\alpha v_\beta)\tilde{W}_1\nonumber\\
&&-g^{\mu\nu}\tilde{W}_2+v^\mu v^\nu \tilde{W}_3+(n^\mu v^\nu+n^\nu v^\mu)\tilde{W}_4
+n^\mu n^\nu\tilde{W}_5.
\end{eqnarray}
The triple differential decay rate can be written in terms of
$\tilde{W}_1,\dots,\tilde{W}_5$ as \cite{Lange:2005yw}

\begin{eqnarray}
\label{eq:dgu}
   &&\frac{d^3\Gamma}{dP_+\,dP_-\,dP_l}
   = \frac{G_F^2|V_{ub}|^2}{16\pi^3}\,(M_B-P_+)\,
    \bigg[(P_- -P_l)(M_B-P_- +P_l-P_+)\,\tilde{W}_1  \nonumber\\
   &+& (M_B-P_-)(P_- -P_+)\,\frac{\tilde{W}_2}{2}
+ (P_- -P_l)(P_l-P_+)\left(\frac y 4 \tilde{W}_3+\tilde{W}_4 +\frac 1 y \tilde{W}_5\right)
   \bigg] \,,
\end{eqnarray}
where
\begin{equation}
\label{eq:ydef}
   y = \frac{P_- -P_+}{M_B-P_+}\,.
\end{equation}

\paragraph{Known \boldmath $1/m_b$ corrections}

The hadronic tensor can be factorized as in equation (\ref{eq:nlope}).
In this paper we will be interested in the terms which are suppressed
by one power of $m_b$.  There are currently two types of these terms
which are known. The first type are ``subleading shape functions" i.e.
properly factorized terms of the form $h^{0}\cdot j^{0}\otimes s^{1}$,
where $h^{0}$ and $j^{0}$ are explicitly known at ${\cal
  O}(\alpha_s^0)$. The second type are ``kinematical power
corrections", i.e. terms \emph{calculated within the parton model}
which are suppressed both by $\alpha_s$ and $1/m_b$. These terms will
be properly factorized in this paper. We now briefly review these two
types.

The contributions of the form $h^{0}\cdot j^{0}\otimes s^{1}$ to the
hadronic tensor, i.e. the subleading shape functions, were calculated
using SCET in \cite{Lee:2004ja,Bosch:2004cb,Beneke:2004in} (for
earlier partial calculations see
\cite{Bauer:2001mh,Leibovich:2002ys,Bauer:2002yu,Neubert:2002yx,Burrell:2003cf}).
Here we use the results of \cite{Bosch:2004cb}. The above contribution
to the hadronic tensor is:
\begin{eqnarray}
\label{eq:Wij_SSF}
   W_{ij}^{\rm \,SSF} &=& \int d\omega\,\delta(n\cdot p+\omega) \left[
    \frac{\omega\,S(\omega) + t(\omega)}{m_b}\,T_2
    + \frac{s(\omega)}{m_b}\,T_1
    + \frac{t(\omega)}{\bar n\cdot p}\,T_3
    + \frac{u(\omega)}{\bar n\cdot p}\,T_1
    - \frac{v(\omega)}{\bar n\cdot p}\,T_4 \right]\nonumber\\
&-&\pi\alpha_s \int d\omega\,\delta(n\cdot p+\omega)
   \left[ \frac{f_u(\omega)}{\bar n\cdot p}\,T_1
   + \frac{f_v(\omega)}{\bar n\cdot p}\,T_4 \right]
\end{eqnarray}
where
\begin{eqnarray}
\label{eq:traces}
   T_1 &=& \frac14\,\mbox{tr}\left[ \Gamma_i\,\nslash\,\Gamma_j\,
    \frac{1+\vslash}{2} \right] , \hspace{1.05cm}
   T_3 = \frac14\,\mbox{tr}\left[ \Gamma_i\,\gamma_\rho^\perp\gamma_5\,
    \Gamma_j\,\frac{1+\vslash}{2}\,\gamma_\perp^\rho\gamma_5 \right] ,
    \nonumber\\
   T_2 &=& \frac18\,\mbox{tr}\,\Big[ \Gamma_i\,\nslash\,\Gamma_j\,
    (\vslash-\nslash) \Big] \,, \qquad
   T_4 = \frac14\,\mbox{tr}\left[ \Gamma_i\,\nslash\gamma_5\,\Gamma_j\,
    \frac{1+\vslash}{2}\,(\vslash-\nslash)\,\gamma_5 \right] .
\end{eqnarray}
The subleading shape functions are defined as:
 \begin{eqnarray}\label{ssf}
\langle\bar h(0)\,[0,x_-]\, h(x_-)\rangle
   &=& \int d\omega\,e^{-\frac{i}{2}\omega\bar n\cdot x}\,S(\omega) \,,
    \nonumber\\[-0.15cm]
m_b\,\langle i\int d^4z\,T\{ \bar h(0)\,[0,x_-]\, h(x_-)
   {\cal L}_h^{(2)}(z) \} \rangle
   &=&  \int d\omega\,e^{-\frac{i}{2}\omega\bar n\cdot x}\,
    s(\omega) \,, \nonumber\\
   \langle\bar h(0)\,\nslash\,[0,x_-]\,(i\Dslash_\perp h)(x_-)\rangle
   &=& \int d\omega\,e^{-\frac{i}{2}\omega\bar n\cdot x}\,t(\omega) \,,
    \nonumber\\[-0.15cm]
   -i\int\limits_0^{\bar n\cdot x/2}\!dt\,
   \langle\bar h(0)\,[0,tn]\,(iD_\perp)^2(tn)\,[tn,x_-]\,h(x_-)\rangle
   &=& \int d\omega\,e^{-\frac{i}{2}\omega\bar n\cdot x}\,u(\omega) \,,
    \nonumber\\[-0.2cm]
   -i\int\limits_0^{\bar n\cdot x/2}\!dt\,
   \langle\bar h(0)\,\frac{\nslash}{2}\,[0,tn]\,
   \sigma_{\mu\nu}^\perp\,gG_\perp^{\mu\nu}(tn)\,[tn,x_-]\,h(x_-)\rangle
   &=& \int d\omega\,e^{-\frac{i}{2}\omega\bar n\cdot x}\,v(\omega) \,,
\end{eqnarray}
and
\begin{eqnarray}\label{4qdef}
   2(-i)^2 &&\hspace{-0.6cm} \int\limits_0^{\bar n\cdot x/2}\!dt_1
   \int\limits_{t_1}^{\bar n\cdot x/2}\!dt_2\,
   \langle \big[ \left( \bar h S\right)_{0} t_a \big]_k
   \big[ t_a \left( S^\dagger h \right)_{x_-} \big]_l
   \big[ \left( \bar q S\right)_{t_2 n} \big]_l\,\nslash
   \big[ \left( S^\dagger q \right)_{t_1 n} \big]_k \rangle \nonumber\\
   &=& \int d\omega\,e^{-\frac{i}{2}\omega\bar n\cdot x}\,f_u(\omega) \,,
    \nonumber\\
   2(-i)^2 &&\hspace{-0.6cm} \int\limits_0^{\bar n\cdot x/2}\!dt_1
   \int\limits_{t_1}^{\bar n\cdot x/2}\!dt_2\,
   \langle \big[ \left( \bar h S\right)_{0} t_a \big]_k\,\nslash\gamma_5
   \big[ t_a \left( S^\dagger h \right)_{x_-} \big]_l
   \big[ \left( \bar q S\right)_{t_2 n} \big]_l\,\nslash\gamma_5
   \big[ \left( S^\dagger q \right)_{t_1 n} \big]_k \rangle \nonumber\\
   &=& \int d\omega\,e^{-\frac{i}{2}\omega\bar n\cdot x}\,f_v(\omega) \,,
\end{eqnarray}
where $k$, $l$ are color indices, $S$ in equation (\ref{4qdef}) is a
soft Wilson line defined in \cite{Bosch:2004cb} (not to be confused
with the leading order shape function $S(\omega)$!), $[x,y]\equiv
S(x)S^\dagger(y)$, ${\cal L}_h^{(2)}$ is the next-to-leading term in
the expansion of the HQET Lagrangian, and
$$
   \langle\bar h\dots h\rangle
   \equiv \frac{\langle\bar B(v)|\,\bar h\dots h\,|\bar B(v)\rangle}{2M_B}.
$$
The second type of terms, namely the kinematical power corrections,
can be found in \cite{Lange:2005yw}, where they were called ${\cal
  F}^{\rm kin}$. In that paper the corrections were convoluted with
the ``tree level shape function", in absence of proper factorization.
The relevant expressions are, for the $Q_{7\gamma}-Q_{7\gamma}$
contribution to $\bar B \to X_s\, \gamma$,
\begin{equation}\label{eq:Wii}
W=-\frac{2}{M_B-P_+}\frac{C_F\,\alpha_s(\bar \mu)}{4\pi}\int _0^{P_+}\,d\hat{\omega}\,\hat {S}(\hat\omega,\mu_i)\left(-15-16\,\ln x\right),
\end{equation}
and for $\bar B \to X_u\,
l\,\bar\nu$,
\begin{eqnarray}\label{eq:kinNLO}
   \tilde{W}_1^{\rm kin(1)}
   &=& \frac{1}{M_B-P_+}\,\frac{C_F\alpha_s(\bar\mu)}{4\pi}
    \int_0^{P_+}\!d\hat\omega\,\hat S(\hat\omega,\mu_i)
    \left[ 6 - \frac5y + \left( \frac{12}{y}-4 \right) \ln\frac{y}{x} \right]
    , \nonumber\\
   \tilde{W}_2^{\rm kin(1)}
   &=& \frac{1}{M_B-P_+}\,\frac{C_F\alpha_s(\bar\mu)}{4\pi}
    \int_0^{P_+}\!d\hat\omega\,\hat S(\hat\omega,\mu_i)
    \left[ \frac2y \right] , \nonumber\\
   \left(\frac{y}{4}\tilde{W}_3+\tilde{W}_4+\frac1{y}\tilde{W}_5\right)^{\rm kin(1)}
   &=& \frac{1}{M_B-P_+}\,\frac{C_F\alpha_s(\bar\mu)}{4\pi}
    \int_0^{P_+}\!d\hat\omega\,\hat S(\hat\omega,\mu_i)
    \left[ 4 - \frac{22}{y} + \frac8y \ln\frac{y}{x} \right] .
\end{eqnarray}
where
$$x=\frac{P_+-\hat\omega}{M_B-P_+}.$$
The ``hatted" function $\hat S(\hat\omega,\mu_i)$ is related to
$S(\omega)$, defined in equation (\ref{ssf}), by a change of variables
and a $1/m_b$ suppressed term. The exact relation can be found in
\cite{Lange:2005yw}.

Since the expansion in \cite{Lange:2005yw} was organized in inverse
powers of $M_B-P_+$ instead of $m_b$, for future reference we will
need also the leading order term for $\tilde W_1$
\begin{equation}\label{eq:BuLP}
    \tilde{W}_1^{\rm (0)}
   = H_{1}(y,\mu_h) \int_0^{P_+}\!d\hat\omega\,y m_b\,
    J(y m_b(P_+-\hat\omega),\mu_i)\,\hat S(\hat\omega,\mu_i) \,,
\end{equation}
where
\begin{eqnarray}
   H_{u1}(y,\mu_h)
   &=& 1 +\! \frac{C_F\alpha_s(\mu_h)}{4\pi} \!\left[\!
    - 4\ln^2\!\frac{y m_b}{\mu_h} + 10\ln\frac{y m_b}{\mu_h} - 4\ln y
    - \frac{2\ln y}{1-y} - 4L_2(1\!-\!y) - \frac{\pi^2}{6} -\! 12 \right]\! ,
    \nonumber\\
\end{eqnarray}
and
\begin{equation}\label{eq:LOjet}
   J(p^2,\mu) = \delta(p^2) \left[ 1
   + \frac{C_F \alpha_s(\mu)}{4\pi}(7-\pi^2) \right]
   + \frac{C_F \alpha_s(\mu)}{4\pi} \left[ \frac{1}{p^2}
   \left( 4\ln \frac{p^2}{\mu^2} - 3\right) \right]_*^{[\mu^2]} .
\end{equation}

\paragraph{Some Ingredients of SCET}

SCET is the appropriate effective field theory to discuss inclusive
$B$ decays in the end point region.  The SCET expansion parameter is
$\sqrt\lambda\equiv\sqrt{\Lambda_{\rm QCD}/m_b}$. For the following we
will need the expansion of the SCET heavy-to-light currents to second
order, as well as the leading order hard-collinear
Lagrangian\footnote{The power corrections to the hard-collinear
  Lagrangian always involve extra soft particles (apart from the heavy
  quark) and as such do not contribute to the subleading jet
  functions.}. These are most conveniently listed in
\cite{Bosch:2004cb}.

First, the hard-collinear Lagrangian is
\begin{eqnarray}
\label{eq:colllag}
   {\cal L}_\xi^{(0)} &=& \bar\xi^{(0)}\,\frac{\nbslash}{2}
   \left( in\cdot D_{hc}^{(0)}
   + i\Dslash_{\perp hc}^{(0)}\,\frac{1}{i\bar n\cdot D_{hc}^{(0)}}\, i\Dslash_{\perp hc}^{(0)} \right) \xi^{(0)} \nonumber\\
 &=& \bar\xi^{(0)}\,\frac{\nbslash}{2}
   \left( in\cdot D_{hc}^{(0)}
   + i\Dslash_{\perp hc}^{(0)}\,W\,\frac{1}{i\bar n\cdot \partial}\,W^\dagger\,
   i\Dslash_{\perp hc}^{(0)} \right) \xi^{(0)} \,,
\end{eqnarray}
where the Lagrangian is written in terms of ``sterile" fields, i.e.
after the soft degrees of freedom were decoupled via a field redefinition. In the last equation
$iD^{(0)\mu}_{hc}=i\partial^\mu+gA^{(0)\mu}_{hc}$ and
\begin{equation}
   W = P\exp\Bigg( ig \int\limits_{-\infty}^0\!dt\,
   \bar n\cdot A_{hc}^{(0)}(x+t\bar n) \Bigg).
\end{equation}

Next we need the expressions for the currents. We present them in
terms of the ``calligraphic fields" \cite{Hill:2002vw}
\begin{equation}
   \X = W^\dagger \xi^{(0)} \,, \qquad
   \A_{hc}^\mu = W^\dagger (iD_{hc}^{(0)\mu} W).
\end{equation}

Generally speaking, in the SCET expansion of the currents (and the
Lagrangian), the power suppression arises from two separate sources,
not mutually exclusive: presence of power suppressed components of the
hard-collinear gluon field or the hard-collinear covariant derivative,
and presence of soft fields or their covariant derivatives.  For the
purpose of this paper we need only currents of the first type, which
are
\begin{eqnarray}\label{eq:SCET_currents}
   J^{(0)} &=& \bar\X\,\Gamma \left( S^\dagger h \right)_{x_-} ,
    \nonumber\\
   J^{(1)} &=&\mbox{}- \bar\X\,\frac{\nbslash}{2}\,\calAslash_{\perp hc}\,
    \frac{1}{i\bar n\cdot\overleftarrow{\partial}}\,\Gamma
    \left( S^\dagger h \right)_{x_-}
    - \bar\X\,\Gamma\,\frac{\nslash}{2m_b}\,
    \calAslash_{\perp hc} \left( S^\dagger h \right)_{x_-} , \nonumber\\
   J^{(2)} &=&\mbox{}- \bar\X\,\Gamma
     \frac{\nslash}{2m_b}
    \, n\cdot\A_{hc} \left( S^\dagger h \right)_{x_-} - \bar\X\,\Gamma\,\frac{1}{i\bar n\cdot\partial}\, n\cdot\A_{hc} \left( S^\dagger h \right)_{x_-}\nonumber\\
   &&\mbox{}- \bar\X\,\Gamma\,\frac{1}{i\bar n\cdot\partial}\,
    \frac{(i\calDslash_{\perp hc}\,\calAslash_{\perp hc})}{m_b}
    \left( S^\dagger h \right)_{x_-}
    + \bar\X\,\frac{i\overleftarrow{\calDslash\!}_{\perp hc}}{m_b}\,
    \frac{1}{i\bar n\cdot\overleftarrow{\partial}}\,\frac{\nbslash}{2}\,
    \Gamma\,\frac{\nslash}{2}\,\calAslash_{\perp hc}
    \left( S^\dagger h \right)_{x_-} \,.
\end{eqnarray}
where we have suppressed the overall $e^{-im_b v\cdot x}$ factor in
each term.  For completeness we list also the currents of the second
type, which we will not use
\begin{eqnarray}
  && J_{\rm not\,  used}^{(1)} = \bar\X\,\Gamma\,x_\perp^\mu
    \left( S^\dagger D_\mu h \right)_{x_-}
    + \bar\X\,\frac{\nbslash}{2}\,i\!\overleftarrow{\delslash}\!_\perp\,
    \frac{1}{i\bar n\cdot\overleftarrow{\partial}}\,\Gamma
    \left( S^\dagger h \right)_{x_-} , \nonumber\\
   &&J_{\rm not\, used}^{(2)} = \bar\X\,\Gamma \left[
    \frac{n\cdot x}{2} \left( S^\dagger \bar n\cdot D h \right)_{x_-}
    + \frac{x_\perp^\mu x_\perp^\nu}{2}
    \left( S^\dagger D_\mu D_\nu h \right)_{x_-}
    + \left( S^\dagger \frac{i\Dslash}{2m_b}\,h \right)_{x_-} \right]
    \nonumber\\
   &+& \bar\X\,\frac{\nbslash}{2}\,
    i\!\overleftarrow{\delslash}\!_\perp\,
    \frac{1}{i\bar n\cdot\overleftarrow{\partial}}\,\Gamma\,x_\perp^\mu
    \left( S^\dagger D_\mu h \right)_{x_-}
- \bar\X \left( \frac{\nbslash}{2}\,\calAslash_{\perp hc}\,
    \frac{1}{i\bar n\cdot\overleftarrow{\partial}}\,\Gamma
    + \Gamma\,\frac{\nslash}{2m_b}\,\calAslash_{\perp hc} \right)
    x_\perp^\mu \left( S^\dagger D_\mu h \right)_{x_-}. \nonumber\\
\end{eqnarray}
The second term in $J_{\rm not\, used}^{(1)}$ does not contain any
soft fields apart from the heavy quark. Still, we can ignore its
contribution in this paper, since we can always set $p_\perp$, where
$p$ is the total hard-collinear momentum, to zero.

It should be noted that these currents were matched at zeroth order in
$\alpha_s(\mu_h)$ and as a result the Wilson coefficients are always
equal to $1$.  When matching beyond zeroth order in $\alpha_s(\mu_h)$,
one would expect more complicated currents with multiple
non-localities, see for example the one loop matching onto the first
order SCET current in \cite{Hill:2004if}. The resulting contributions
to the hadronic tensor would be suppressed by
$\alpha_s(\mu_h)\times\alpha_s(\mu_i)\times 1/m_b$ and as such are
much smaller than the contributions considered in this paper.

\section{Analysis by regions}
\label{sec:regions} In this section we will calculate, within the
parton model, the one loop corrections to the hadronic tensor, defined
in equation (\ref{eq:Wij}), that scale as ${\cal O}(\lambda^0)$ in
the shape-function region, i.e.  terms which are suppressed by $1/m_b$
compared to the leading order terms, which scale as ${\cal
  O}(\lambda^{-1})$. We perform the calculation for a general Dirac
structure in the hadronic tensor. In order to simplify the calculation
we neglect the residual momentum of the $b$ quark, i.e. we work with
on-shell $b$ quarks. As a result, all the terms are constants or
logarithms of the form $\ln\, (n\cdot p)/(\bar n\cdot p)\equiv\ln\,
r$, where $p$ is the partonic momentum of the jet, defined in section
\ref{sec:review}, which satisfies $p_\perp=0$ and $p^2=\bar{n}\cdot
p\,\,n\cdot p$.

\begin{figure}
\begin{center}
\begin{tabular}{ccccccc}
\epsfig{file=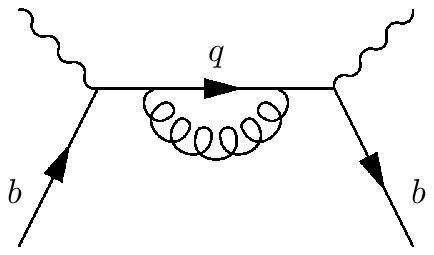,width=4cm}&&&&&&
\epsfig{file=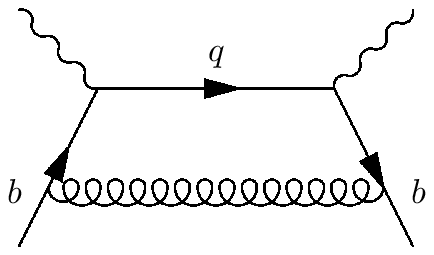,width=4cm}\\\\\\
\epsfig{file=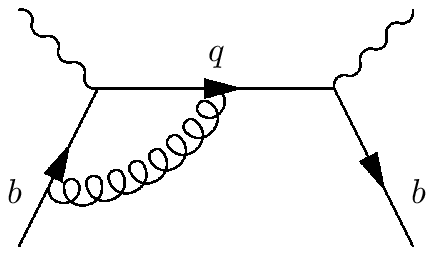,width=4cm}&&&&&&
\epsfig{file=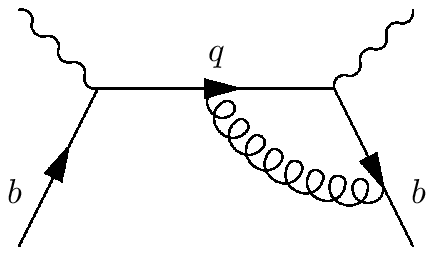,width=4cm}
\end{tabular}
\caption{\label{fig:hadten} One loop diagrams contributing to the
  hadronic tensor, top left: ``Self energy" diagram, top right: ``Box"
  diagram, bottom line: two ``Vertex" diagrams. The letter next to
  each solid line denotes the flavor of the quark.}
\end{center}
\end{figure}

We perform the calculation both in ``full QCD" and by using the method
or regions \cite{Beneke:1997zp, Smirnov:2002pj}. We find that only two
kinematical regions are needed: a hard-collinear region, where the
loop momentum scales as $(1,\lambda,\lambda^{1/2})$, and a soft region,
where the loop momentum scales as $(\lambda,\lambda,\lambda)$. There is
no contribution from a hard region, were the loop momentum scales as
$(1,1,1)$, which is in accordance with the lack of terms of the form
$h^{1}\cdot j^{0}\otimes s^{0}$ in the factorization formula. As
explained in the introduction this is a general result which holds
also beyond one loop order. Typically, we find that terms of the form
$\ln r$ can be decomposed as:
\begin{equation}\label{eq:lnr}
  \ln r\equiv \ln \frac{n\cdot p}{\bar n\cdot p}=\underbrace{-\frac{1}{\epsilon}+\ln \frac{\mu^2}{p^2}}_{\rm hard-collinear}+
\underbrace{\frac{1}{\epsilon}+\ln \frac{(n\cdot p)^2}{\mu^2}}_{\rm soft}.
\end{equation}
At one loop there are several diagrams that contribute to the hadronic
tensor. These diagrams are shown in figure \ref{fig:hadten}.  We
present the results for each diagram separately, namely, the ``Self
energy" diagram (top left), the ``Box" diagram (top right), and the
``Vertex" contribution which is the {\em sum} of the diagrams on the
bottom line of figure \ref{fig:hadten}.  We use Feynman gauge
throughout this paper.

\subsection{Full QCD}
Calculating the diagrams we find the following results:
\begin{itemize}
\item[]``Self energy":
\begin{equation}
\label{eq:QCDSE}
W_{ij}=\frac{C_F\alpha_s}{4\pi}\,\theta(p^2)\,\mbox{tr}\left[\Gamma_i\,\nbslash\,\Gamma_j\,\frac{1+\vslash}{2}\right]\,\frac1{\bar n\cdot p}\,\frac14,
\end{equation}
\item[]``Box"
  \begin{eqnarray}
\label{eq:QCDBox}
W_{ij}=\frac{C_F\alpha_s}{4\pi}\,\theta(p^2)&\Bigg\{&\mbox{tr}\left[\Gamma_i\,\nslash\,\Gamma_j\,\frac{1+\vslash}{2}\right]\,\frac1{\bar n\cdot p}\,
\left(-1-\ln r\right)+ \mbox{tr}\left[\Gamma_i\,\nslash\,\Gamma_j\,\nbslash\right]\,\frac1{m_b}\,\frac14\nonumber\\
&+&\mbox{tr}\left[\Gamma_i\,\nslash\,\Gamma_j\,\nslash\right]\frac1{m_b}\,\frac14\left(-2-\ln r\right)+\mbox{tr}\left[\Gamma_i\,\nslash\,\Gamma_j\,\nslash\right]\,\frac{\bar n \cdot p}{m_b^2}\,\frac{1}{16}\Bigg\},
\end{eqnarray}
\item[]``Vertex"
\begin{eqnarray}
\label{eq:QCDVer}
W_{ij}&=&\frac{C_F\alpha_s}{4\pi}\,\theta(p^2)\Bigg\{\mbox{tr}\left[\Gamma_i\,\nslash\,\Gamma_j\,\nbslash\right]\,\frac1{m_b}\,\frac14\left(\frac32+\ln r\right)-\mbox{tr}\left[\Gamma_i\,\nslash\,\Gamma_j\,\frac{1+\vslash}{2}\right]\,\frac1{\bar n\cdot p}\,
\left(1+2\ln r\right)\nonumber\\
&-&\left(\mbox{tr}\left[\Gamma_i\,\nslash\,\gamma^\beta\,\gamma_\perp^\alpha\,\Gamma_j\,\gamma^\perp_\alpha\,\gamma_\beta\,\frac{1+\vslash}{2}\right]+\mbox{tr}\left[\Gamma_i\,\gamma_\perp^\alpha\,\gamma^\beta\,\nslash\,\Gamma_j\,\frac{1+\vslash}{2}\,\gamma_\beta\,\gamma^\perp_\alpha\right]\right)\,\frac{1}{m_b}\,\frac{1}{16}\nonumber\\
&-&\mbox{tr}\left[\Gamma_i\,\gamma_\perp^\alpha\,\Gamma_j\,\gamma^\perp_\alpha\right]\frac1{m_b}\,\frac18\Bigg\}.
\end{eqnarray}
\end{itemize}
In order to check these results, we can compare them to the expansion
of the one loop expressions of the hadronic tensor for $\bar B \to X_u\,
l\,\bar\nu$, and the $Q_{7\gamma}-Q_{7\gamma}$ contribution to $\bar B \to
X_s\, \gamma$. This is most easily done by using equations
(\ref{eq:Wii})-(\ref{eq:LOjet}) in section \ref{sec:review}, which are
taken from \cite{Lange:2005yw}. In that paper the correction were
convoluted with the ``tree level shape function".  We can undo this
convolution by the replacements:
\begin{equation}
M_B-P_+\to m_b,\quad y\to \frac{\bar n\cdot p}{m_b}, \quad \frac{x}{y}\to \frac {n\cdot p}{\bar n \cdot p},\quad \int _0^{P_+}\,d\hat{\omega}\,\hat {S}(\hat\omega,\mu_i)\to 1.
\end{equation}
For $\bar B \to X_u\, l\,\bar\nu$ we also need to expand $\tilde{W}_1^{\rm (0)}$ in
powers of ${n\cdot p}/{\bar n \cdot p}$. In total we find for
$\bar B \to X_s\, \gamma$:
\begin{equation}
\label{eq:QCDb2s}
W=-\frac{C_F\alpha_s}{4\pi}\frac2{m_b}\left(-15-16\,\ln r\right),
\end{equation}
and for $\bar B \to X_u\, l\,\bar\nu$:
\begin{eqnarray}
\label{eq:QCDb2u}
\tilde{W}_1&=&\frac{C_F\alpha_s}{4\pi}\left(\frac{10}{m_b}-\frac{9}{\bar n\cdot p}-\frac{12}{\bar n\cdot p}\ln r+\frac{4}{m_b}\ln r\right)\nonumber\\
\tilde{W}_2&=&\frac{C_F\alpha_s}{4\pi}\frac{2}{\bar n\cdot p}\nonumber\\
\frac{y}{4}\tilde{W}_3+\tilde{W}_4+\frac1{y}\tilde{W}_5&=&\frac{C_F\alpha_s}{4\pi}\left(\frac{4}{m_b}-\frac{22}{\bar n\cdot p}-\frac8{\bar n\cdot p}\ln r\right).
\end{eqnarray}
Summing over (\ref{eq:QCDSE}), (\ref{eq:QCDBox}), and
(\ref{eq:QCDVer}), and calculating the traces for each decay mode
using the expressions after equation (\ref{eq:Wij}), we find complete
agreement with (\ref{eq:QCDb2s}) and (\ref{eq:QCDb2u}).

We are now ready to repeat this calculation using the method of
regions. We perform the calculation in $d=4-2\epsilon$ dimensions and
use dimensional regularization to regularize both the UV and IR
divergences. We also implicitly take $\mu\to \mu
e^{\gamma_E/2}(4\pi)^{-1/2}$.

\subsection{Hard-collinear region}
For the hard-collinear region we find the following results.
\begin{itemize}
\item[]``Self energy":
\begin{equation}
\label{eq:SE_coll}
W_{ij}=\theta(p^2)\frac{C_F\alpha_s}{4\pi}\,\mbox{tr}\left[\Gamma_i\,\nbslash\,\Gamma_j\,\frac{1+\vslash}{2}\right]\,\frac1{\bar n\cdot p}\,\frac14,
\end{equation}
\item[]``Box"
\begin{eqnarray}
\label{eq:Box_coll}
W_{ij}&=&\theta(p^2)\frac{C_F\alpha_s}{4\pi}\Bigg\{\mbox{tr}\left[\Gamma_i\,\nslash\,\Gamma_j\,\frac{1+\vslash}{2}\right]\,\frac1{\bar n\cdot p}\,
\left(-\frac1{\epsilon}+2-\ln\frac {\mu^2}{p^2}\right)\nonumber\\
&+&\mbox{tr}\left[\Gamma_i\,\nslash\,\Gamma_j\,\nslash\right]\frac1{m_b}\,\frac14\,\left(-\frac1{\epsilon}-1-\ln\frac {\mu^2}{p^2}\right)+\mbox{tr}\left[\Gamma_i\,\nslash\,\Gamma_j\,\nslash\right]\,\frac{\bar n \cdot p}{m_b^2}\,\frac{1}{16}\Bigg\},
\end{eqnarray}
\item[]``Vertex"
\begin{eqnarray}
\label{eq:Ver_coll}
W_{ij}&=&\theta(p^2)\frac{C_F\alpha_s}{4\pi}\Bigg\{\mbox{tr}\left[\Gamma_i\,\nslash\,\Gamma_j\,\frac{1+\vslash}{2}\right]\,\frac{2}{\bar n\cdot p}\,
\left(-\frac1{\epsilon}+\frac12-\ln\frac {\mu^2}{p^2}\right)+\nonumber\\
&+&\mbox{tr}\left[\Gamma_i\,\nslash\,\Gamma_j\,\nbslash\right]\,\frac1{m_b}\,\frac14\left(\frac1{\epsilon}+\frac32+\ln\frac {\mu^2}{p^2}\right)-\mbox{tr}\left[\Gamma_i\,\gamma_\perp^\alpha\,\Gamma_j\,\gamma^\perp_\alpha\right]\frac1{m_b}\,\frac18\nonumber\\
&-&\left(\mbox{tr}\left[\Gamma_i\,\nslash\,\gamma^\beta\,\gamma_\perp^\alpha\,\Gamma_j\,\gamma^\perp_\alpha\,\gamma_\beta\,\frac{1+\vslash}{2}\right]+\mbox{tr}\left[\Gamma_i\,\gamma_\perp^\alpha\,\gamma^\beta\,\nslash\,\Gamma_j\,\frac{1+\vslash}{2}\,\gamma_\beta\,\gamma^\perp_\alpha\right]\right)\,\frac{1}{m_b}\,\frac{1}{16}\Bigg\}.\nonumber\\
\end{eqnarray}
\end{itemize}
\subsection{Soft Region}
For the soft region we find that the ``self energy" diagram does not contribute. For the other diagrams we have:
\begin{itemize}
\item[] ``Box"
\begin{eqnarray}\label{eq:softbox}
W_{ij}&=&\theta(n\cdot p)\frac{C_F\alpha_s}{4\pi}\Bigg\{\mbox{tr}\left[\Gamma_i\,\nslash\,\Gamma_j\,\frac{1+\vslash}{2}\right]\,\frac1{\bar n\cdot p}\,
\left(\frac1{\epsilon}-3-\ln\frac {(n\cdot p)^2}{\mu^2}\right)\nonumber\\
&+& \mbox{tr}\left[\Gamma_i\,\nslash\,\Gamma_j\,\nbslash\right]\,\frac1{m_b}\,\frac14+\mbox{tr}\left[\Gamma_i\,\nslash\,\Gamma_j\,\nslash\right]\frac1{m_b}\,\frac14\left(\frac1{\epsilon}-1-\ln\frac {(n\cdot p)^2}{\mu^2}\right)\Bigg\},
\end{eqnarray}
\item[]``Vertex"
\begin{eqnarray}\label{eq:softvertex}
W_{ij}&=&\theta(n\cdot p)\frac{C_F\alpha_s}{4\pi}\Bigg\{\mbox{tr}\left[\Gamma_i\,\nslash\,\Gamma_j\,\frac{1+\vslash}{2}\right]\,\frac1{\bar n\cdot p}\,2
\left(\frac1{\epsilon}-1-\ln\frac {(n\cdot p)^2}{\mu^2}\right)+\nonumber\\
&+& \mbox{tr}\left[\Gamma_i\,\nslash\,\Gamma_j\,\nbslash\right]\,\frac1{m_b}\,\frac14\left(-\frac1{\epsilon}+\ln\frac {(n\cdot p)^2}{\mu^2}\right)\Bigg\}
\end{eqnarray}
\end{itemize}
Adding up the two types of regions we find that, as expected, the sum
of the hard-collinear and soft regions reproduce the full QCD result.
Notice also that the structure of the soft region is simpler than that
of the hard-collinear region. In the next section we will see that the
reason is that the soft region is accounted for by the parton level
one loop expressions for only two subleading shape function, while for
the hard-collinear region we need several subleading jet functions.
\section{Effective field theory  calculation}\label{sec:EFT}
We have seen in the previous section that the one loop corrections to
the hadronic tensor which scale as ${\cal O}(\lambda^0)$ in the end
point region, arise from two kinematical regions: a hard-collinear
region and a soft region. In this section we will see how the soft
region is accounted for by the one loop corrections to the subleading
shape functions' contribution calculated within the parton model.  The
hard-collinear region is accounted for by the contribution of the time
ordered product of power suppressed SCET currents. This calculation is
the basis for the subleading jet function calculation which we perform
in the next section.
\subsection{Soft contribution}
\begin{figure}
\begin{center}
\begin{tabular}{ccc}
\epsfig{file=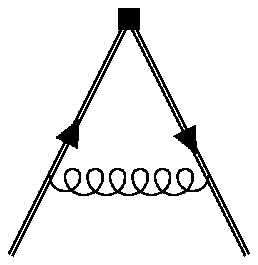,width=2cm}&
\epsfig{file=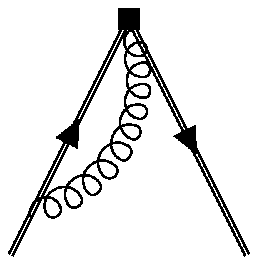,width=2cm}&
\epsfig{file=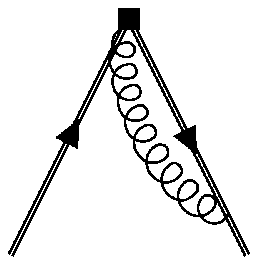,width=2cm}
\end{tabular}
\caption{\label{fig:udiag} One loop diagrams contributing to the parton level expression of $u(\omega)$. }
\end{center}
\end{figure}

The contribution of the soft region can be fully accounted for by
calculating, within the parton model, the one loop corrections to the
subleading shape functions. In particular there is no need to
introduce new subleading shape functions.  From equation
(\ref{eq:Wij_SSF}) we see that the hadronic tensor depends on several
subleading shape functions. Naively, one would assume that we need to
calculate the one loop corrections for $\omega S(\omega), s(\omega),
t(\omega), u(\omega), v(\omega)$, as well as the four-quark shape
functions $f_u(\omega), f_v(\omega)$. In practice, only the
contributions of $\omega S(\omega)$ and $ u(\omega)$ are needed for the
following reasons.
\begin{itemize}
\item[-] We have chosen the coordinate system such that $v_\perp=0$. As a result the matrix elements of the operator corresponding to $t(\omega)$ and $v(\omega)$,
vanish at one loop, since they only contain gluons which have perpendicular polarization.
\item[-] The matrix elements of the operators corresponding to $f_u(\omega)$ and $f_v(\omega)$ vanish since they involve scaleless integrals over the $\bar n$ component of the light quark momentum.
\item[-] Setting the residual momentum of the heavy quarks to zero, we find that the matrix element
of the operator corresponding to $s(\omega)$ vanishes.
\end{itemize}
As a result we need the ``zero external gluon" matrix elements of the operators corresponding to $\omega S(\omega)$ and $u(\omega)$.
The first can be extracted
from \cite{Bosch:2004th}. After setting the residual momentum $k$ to zero, we find
\begin{equation}
\label{eq:wS}
\omega\,S^{\rm parton}_{\rm bare}=\theta(-\omega)\,\frac{C_F\alpha_s}{\pi}\,\left(-\frac1{\epsilon}+\ln\frac{\omega^2}{\mu^2}+1\right).
\end{equation}
For $u(\omega)$, we need to calculate the one loop amplitude which is the sum of the diagrams shown in figure \ref{fig:udiag}.
The relevant Feynman rules needed for this calculation involve zero and one external gluon. We have
\begin{eqnarray}
\vcenter{\epsfig{file=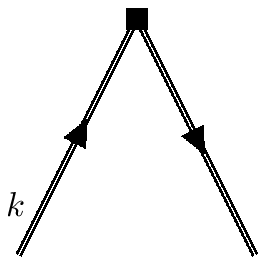,width=2cm}} \hspace{-32em}&&-k_\perp^2\,\delta(\omega-n\cdot k)\nonumber\\
\vcenter{\epsfig{file=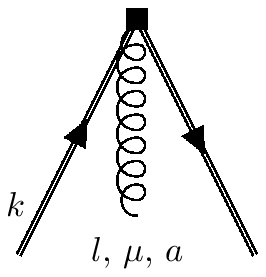,width=2cm}}\hspace{-32em}&&\frac{t^a\,g\,n^\mu}{n\cdot l}\Big[(k-l)^2_\perp\,\delta'(\omega-n\cdot k+n\cdot l)-k^2_\perp\,\delta'(\omega-n\cdot k)\Big]\nonumber\\
&+&\frac{t^a\,g\,n^\mu}{(n\cdot l)^2}(2k_\perp\cdot l_\perp-l^2_\perp)\,\Big[\delta(\omega-n\cdot k+n\cdot l)-\delta(\omega-n\cdot k)\Big].
\end{eqnarray}
The notation is such that $k$ is the incoming heavy quark residual momentum and $l,\,\mu,\,a$ are the outgoing gluon's momentum, polarization, and color index, respectively.  For the one gluon Feynman rule we have omitted terms in which the gluon has a perpendicular polarization, since such terms do not contribute to the one loop amplitude for $v_\perp=0$.

These Feynman rules were first calculated by Trott and Williamson in
\cite{Trott:2005vw}, where the corresponding operator is called
$Q_3(\omega,\Gamma)$.  Keeping only the terms which are
proportional to $n^\mu$ in the one external gluon Feynman rule of
\cite{Trott:2005vw}, our result agrees with theirs accounting for an
overall factor of $2/m_b$ and a different sign of the heavy quark
residual momentum, both arising from the slightly different definition
of $u(\omega)$.  Calculating the one loop amplitude, which was not
calculated in \cite{Trott:2005vw}, and setting the residual momentum
$k$ to zero, we find
\begin{equation}
\label{eq:u}
u^{\rm parton}_{\rm bare}=\theta(-\omega)\,\frac{C_F\alpha_s}{\pi}\,\left(\frac3{\epsilon}-3\ln\frac{\omega^2}{\mu^2}-5\right).
\end{equation}
Inserting (\ref{eq:wS}) and (\ref{eq:u}) into (\ref{eq:Wij_SSF}), and
setting $s(\omega),\,t(\omega),\, v(\omega),\, f_u(\omega)$ and $f_v(\omega)$ to zero, we find:
\begin{eqnarray}
  W_{ij}^{(2)} = \theta(n\cdot p) \frac{C_F\alpha_s}{4\pi}&\Bigg\{&\mbox{tr}\left[\Gamma_i\,\nslash\,\Gamma_j\,\frac{1+\vslash}{2}\right]\,\frac1{\bar n\cdot p}\,\left(\frac3{\epsilon}-3\ln\frac{(n \cdot p)^2}{\mu^2}-5\right)\nonumber\\
  &+& \mbox{tr}\left[\Gamma_i\,\nslash\,\Gamma_j\,(\nbslash-\nslash)\right]\,\frac1{m_b}\,\frac14\left(-\frac1{\epsilon}+\ln\frac{(n \cdot p)^2}{\mu^2}+1\right)\Bigg\},
\end{eqnarray}
which is the total contribution of the soft region, i.e. the sum of equations (\ref{eq:softbox}) and (\ref{eq:softvertex}).

At this point we should note that the question of operator mixing and
renormalization with regard to $u(\omega)$ is still open, as it was
not considered in \cite{Trott:2005vw}. The main complication being the
need to introduce new subleading shape functions which $u(\omega)$ can
mix into, and establishing the closure of the basis. The analysis of
this question goes beyond the scope of this paper. For our purposes
the important point is that the terms of the form $h^{0}\cdot
j^{0}\otimes s^{1}$, that were already calculated in the literature,
reproduce the contribution of the soft region. As a result in the
final factorization formula, no new terms are needed to account for
this contribution.

\subsection{Hard-Collinear contribution}\label{sec:hc}
In order to reproduce the result of the hard-collinear region, we need
to consider various combinations of the SCET currents and the SCET
Lagrangian\footnote{It is easy to show that the TOP of a first order
  current with the zeroth order current vanish since the perpendicular
  components of the hard-collinear momentum can be chosen to be zero.
  This corresponds of course to the absence of $1/\sqrt{m_b}$
  corrections to the hadronic tensor.}.  Symbolically we need the
following combinations:
\begin{equation}
J^{\dagger(1)}\,J^{(1)},\quad J^{\dagger(2)}\,J^{(0)}+J^{\dagger(0)}\,J^{(2)}, \quad J^{\dagger(2)}\,J^{(0)}\int d^4 z \,{\cal L}_\xi^{(0)}+J^{\dagger(0)}\,J^{(2)}\int d^4 z \,{\cal L}_\xi^{(0)},
\end{equation}
where ${\cal L}_\xi^{(0)}$ and $J^{(i)}$ are defined in equations
(\ref{eq:colllag}) and (\ref{eq:SCET_currents}), respectively.
Calculating the contribution of each combination we find that the non
zero combinations are:
\begin{itemize}
\item  $J^{\dagger(1)}\,J^{(1)}$
\begin{eqnarray}
\label{eq:collfirst}
W_{ij}&=&\theta(p^2)\frac{C_F\alpha_s}{4\pi}\,\Bigg\{\mbox{tr}\left[\Gamma_i\,\nbslash\,\Gamma_j\,\frac{1+\vslash}{2}\right]\,\frac1{\bar n\cdot p}\,\frac14+\mbox{tr}\left[\Gamma_i\,\nslash\,\Gamma_j\,\nslash\right]\,\frac{\bar n \cdot p}{m_b^2}\,\frac{1}{16}\nonumber\\
&+&\left(\mbox{tr}\left[\Gamma_i\,\nbslash\,\nslash\,\gamma_\perp^\alpha\,\Gamma_j\,\nslash\gamma^\perp_\alpha\,\frac{1+\vslash}{2}\right]+\mbox{tr}\left[\Gamma_i\,\nslash\,\nbslash\,\gamma_\perp^\alpha\,\Gamma_j\,\frac{1+\vslash}{2}\,\gamma^\perp_\alpha\,\nslash\right]\right)\,\frac{1}{m_b}\,\frac{1}{32}\Bigg\}
\end{eqnarray}\item[]
\item  $J^{\dagger(2)}\,J^{(0)}+J^{\dagger(0)}\,J^{(2)},\quad J^{(2)} =\mbox{}\mbox{}- \bar\X\,\Gamma
     \frac{\nslash}{2m_b}
    \, n\cdot\A_{hc} \left( S^\dagger h \right)_{x_-} $
\begin{eqnarray}\label{eq:nA}
W_{ij}=\theta(p^2)\frac{C_F\alpha_s}{4\pi}&\Bigg\{&
\mbox{tr}\left[\Gamma_i\,\nslash\,\Gamma_j\,\nslash\right]\frac1{m_b}\,\frac14\,\left(-\frac1{\epsilon}-1-\ln\frac {\mu^2}{p^2}\right)\Bigg\}
\end{eqnarray}\item[]
\item  $J^{\dagger(2)}\,J^{(0)}+J^{\dagger(0)}\,J^{(2)},\quad J^{(2)} =\mbox{}- \bar\X\,\Gamma\,\frac{1}{i\bar n\cdot\partial}\, n\cdot\A_{hc} \left( S^\dagger h \right)_{x_-} $
\begin{eqnarray}\label{eq:n'A}
W_{ij}=\theta(p^2)\frac{C_F\alpha_s}{4\pi}&\Bigg\{&\mbox{tr}\left[\Gamma_i\,\nslash\,\Gamma_j\,\frac{1+\vslash}{2}\right]\,\frac1{\bar
n\cdot p}\, \left(-\frac1{\epsilon}+2-\ln\frac
{\mu^2}{p^2}\right)\Bigg\}
\end{eqnarray}\item[]
\item  $J^{\dagger(2)}\,J^{(0)}\int d^4 z \,{\cal L}_\xi^{(0)}+J^{\dagger(0)}\,J^{(2)}\int d^4 z \,{\cal L}_\xi^{(0)},\quad J^{(2)} =\mbox{}\mbox{}- \bar\X\,\Gamma
     \frac{\nslash}{2m_b}
    \, n\cdot\A_{hc} \left( S^\dagger h \right)_{x_-} $
\begin{eqnarray}\label{eq:nAl}
W_{ij}=\theta(p^2)\frac{C_F\alpha_s}{4\pi}&\Bigg\{&
\mbox{tr}\left[\Gamma_i\,\nslash\,\Gamma_j\,\nslash\right]\frac1{m_b}\,\frac14\,\left(-\frac1{\epsilon}-\frac32-\ln\frac {\mu^2}{p^2}\right)\Bigg\}
\end{eqnarray}
\item[]\item  $J^{\dagger(2)}\,J^{(0)}\int d^4 z \,{\cal L}_\xi^{(0)}+J^{\dagger(0)}\,J^{(2)}\int d^4 z \,{\cal L}_\xi^{(0)},\quad J^{(2)} =\mbox{}- \bar\X\,\Gamma\,\frac{1}{i\bar n\cdot\partial}\, n\cdot\A_{hc} \left( S^\dagger h \right)_{x_-} $
\begin{eqnarray}\label{eq:n'Al}
W_{ij}=\theta(p^2)\frac{C_F\alpha_s}{4\pi}&\Bigg\{&\mbox{tr}\left[\Gamma_i\,\nslash\,\Gamma_j\,\frac{1+\vslash}{2}\right]\,\frac2{\bar
n\cdot p}\, \left(-\frac1{\epsilon}+\frac12-\ln\frac
{\mu^2}{p^2}\right)\Bigg\}
\end{eqnarray}
\item[]\item  $J^{\dagger(2)}\,J^{(0)}\int d^4 z \,{\cal L}_\xi^{(0)}+J^{\dagger(0)}\,J^{(2)}\int d^4 z \,{\cal L}_\xi^{(0)},\quad J^{(2)} =\mbox{}- \bar\X\,\Gamma\,\frac{1}{i\bar n\cdot\partial}\,
    \frac{(i\calDslash_{\perp hc}\,\calAslash_{\perp hc})}{m_b}
    \left( S^\dagger h \right)_{x_-}$
\begin{eqnarray}
W_{ij}&=&\theta(p^2)\frac{C_F\alpha_s}{4\pi}\Bigg\{
\mbox{tr}\left[\Gamma_i\,\nslash\,\Gamma_j\,\frac{1+\vslash}{2}\right]\frac1{m_b}\,\left(-\frac1{\epsilon}-\frac32-\ln\frac {\mu^2}{p^2}\right)\nonumber\\
&-&\left(\mbox{tr}\left[\Gamma_i\,\nslash\,\gamma_\perp^\beta\,\gamma_\perp^\alpha\,\Gamma_j\,\gamma^\perp_\alpha\,\gamma^\perp_\beta\,\frac{1+\vslash}{2}\right]+\mbox{tr}\left[\Gamma_i\,\gamma_\perp^\alpha\,\gamma_\perp^\beta\,\nslash\,\Gamma_j\,\frac{1+\vslash}{2}\,\gamma^\perp_\beta\,\gamma^\perp_\alpha\,\right]\right)\,\frac{1}{m_b}\,\frac{1}{16}\Bigg\}\nonumber\\
\end{eqnarray}
\item  $J^{\dagger(2)}\,J^{(0)}\int d^4 z \,{\cal L}_\xi^{(0)}+J^{\dagger(0)}\,J^{(2)}\int d^4 z \,{\cal L}_\xi^{(0)},\quad J^{(2)} =\bar\X\,\frac{i\overleftarrow{\calDslash\!}_{\perp hc}}{m_b}\,
    \frac{1}{i\bar n\cdot\overleftarrow{\partial}}\,\frac{\nbslash}{2}\,
    \Gamma\,\frac{\nslash}{2}\,\calAslash_{\perp hc}
    \left( S^\dagger h \right)_{x_-}$
\begin{eqnarray}
\label{eq:colllast}
W_{ij}&=&\theta(p^2)\frac{C_F\alpha_s}{4\pi}\,\Bigg\{\mbox{tr}\left[\Gamma_i\,\nslash\,\nbslash\,\gamma_\perp^\alpha\,\Gamma_j\,\nslash\gamma^\perp_\alpha\,\frac{1+\vslash}{2}\right]+\mbox{tr}\left[\Gamma_i\,\nbslash\,\nslash\,\gamma_\perp^\alpha\,\Gamma_j\,\frac{1+\vslash}{2}\,\gamma^\perp_\alpha\,\nslash\right]\Bigg\}\,\frac{1}{m_b}\,\frac{1}{16}\nonumber\\
\end{eqnarray}
\end{itemize}
Using $\gamma_\perp^\beta=\gamma^\beta-\bar
n^\beta\nslash/2-n^\beta\nbslash/2$ and $2\vslash=\nslash+\nbslash$,
we can show that the sum of (\ref{eq:collfirst})$-$(\ref{eq:colllast})
is equal to the hard-collinear contribution i.e. the sum of
(\ref{eq:SE_coll}),(\ref{eq:Box_coll}), and (\ref{eq:Ver_coll}).
Again we see that the contribution of the hard-collinear region is
more complicated than that of the soft region. As we will see in the
next section there are seven different jet functions that contribute
at one loop, compared to only two subleading shape function that are
needed to reproduce the soft region.

\section{Subleading jet functions}
\label{sec:subjet}
Following the calculations of the previous sections, it is clear that
in order to properly factorize all the terms in the hadronic tensor
that are both $\alpha_s$ and $1/m_b$ suppressed in the end point
region, we have keep the subleading shape functions' contribution and
replace the so called ``$1/m_b$ kinematical corrections" of equations
(\ref{eq:Wii}) and (\ref{eq:kinNLO}) by the contribution of the
subleading jet functions.  By ``jet function" we mean the
discontinuity of a Fourier transform of a vacuum expectation value of
a time ordered product of hard-collinear fields.  By ``subleading" we
mean that these functions scale as ${\cal O}(\lambda^0)$ in the end
point region.

Before going into the details of the analysis of each subleading jet
function, we wish to make some general remarks. The first issue is
factorization. In the following we establish factorization formula for
terms of the form $h^{0}\cdot j^{1}\otimes s^{0}$. More explicitly, we
always have $h^0\equiv 1$ and $s^0$ is the leading order shape
function. The hadronic tensor factorizes as
\begin{equation}
\label{eq:fact}
W_{ij}=\sum_a C_a\,\, \mbox{tr}\left[\,\Gamma_i\,...\,\Gamma_j\,...\right]\int d\omega\, j_a(p_\omega^2)\,S(\omega).
\end{equation}
Here $C_a$ is a simple kinematical factor of mass dimension -1, e.g.
$1/\bar{n}\cdot p,\, 1/m_b$ and the argument of the jet function $j_a$
is $p_\omega^2=\bar{n}\cdot p\,(n\cdot p+\omega)$.  The factorization
formula can be proven in an analogous way to the leading order
factorization proof as presented, for example, in \cite{Bosch:2004th}.
The only difference is that we have subleading jet functions instead
of a leading jet function.

The second issue, which does not arise at leading order, is the
``correct" definition of the subleading jet functions and the role of
parity and time reversal ($PT$) symmetry. In general the subleading
jet function is the discontinuity of TOP of two {\em different}
combinations of hard-collinear fields $O_a$ and $O_b$. Here $O$ can be
a hard-collinear quark or a product of hard-collinear quark and a
hard-collinear gluon, or even a more complicated object. As a result
we typically have both
\begin{equation}
\int d^4x\, e^{-ipx}\, \langle \Omega |T\left\{O_a^\dagger(0),O_b(x) \right\}|\Omega\rangle\quad
\mbox{and}\quad
\int d^4x\, e^{-ipx}\, \langle \Omega |T\left\{O_b^\dagger(0),O_a(x) \right\}|\Omega\rangle.
\end{equation}
The subleading jet function(s) should be the discontinuity of the {\em
  sum} of the two terms. In practice we find that after we decompose
each TOP according to the different color and Lorentz structures, we
can use translation invariance and the $PT$ symmetry of the strong
interaction to relate the two TOPs. We can therefore define the
subleading jet function as the discontinuity of the coefficient of a
specific structure in \emph{either} TOP.

We will illustrate both issues in the more detailed calculation of the
first subleading jet function. These details will be suppressed in the
derivation of subsequent jet functions. As before, the one loop
calculation are performed in $d=4-2\epsilon$ dimensions and we
implicitly take $\mu\to \mu e^{\gamma_E/2}(4\pi)^{-1/2}$.

The following list of the subleading jet functions is not necessarily
the complete list of possible subleading jet functions. For example,
if the QCD currents are matched onto SCET beyond tree level we would
expect more complicated functions which can depend on more than one
variable. We wish to emphasize, though, that the list is sufficient to
describe all the terms of the form of equation (\ref{eq:fact}) at
zeroth order in $\alpha_s(\mu_h)$ and to all orders in
$\alpha_s(\mu_i)$. In particular it includes all the terms of the form
$h^{0}\cdot j^{1}\otimes s^{0}$ at ${\cal O}(\alpha_s)$.

\subsection{The list of subleading jet functions}
\label{subsec:sjfdef}
\subsubsection{\boldmath $j_n$}
This jet function arises from the TOP of the leading order current
$J^{(0)} =e^{-im_b v\cdot x}\,\bar\X\,\Gamma \left( S^\dagger h
\right)_{x_-}$ and the second order current $J^{(2)} =- e^{-im_b
  v\cdot x}\,\bar\X\,\Gamma\frac{\nslash}{2m_b}\, n\cdot\A_{hc} \left(
  S^\dagger h \right)_{x_-}$. These currents are matched at tree level
and as a result the Wilson coefficients equal $1$. Consequently,
the hard function, which is simply the product of the Wilson
coefficients, equals to 1 also. The contribution of this combination
of currents to the hadronic tensor is given by
\begin{eqnarray}
\label{eq:Wij_n}
W_{ij}&=&\frac{1}{2\pi M_B}\,{\rm
Im}\,\langle\bar{B}|\,i\int\,d^4x\, e^{i(q-m_bv)\cdot x}\,
T\left\{\left(\bar h S\right)\,\Gamma_i\,\X(0),
\,- \bar\X\,\Gamma_j\frac{\nslash}{2m_b}\, n\cdot\A_{hc} \left( S^\dagger h \right)_{x_-} \right\}|\bar{B}\rangle+
\nonumber\\
&+&\frac{1}{2\pi M_B}\,{\rm
Im}\,\langle\bar{B}|\,i\int\,d^4x\, e^{i(q-m_bv)\cdot x}\,
T\left\{-\left(\bar h S\right)n\cdot\A_{hc}\,\frac{\nslash}{2m_b}\Gamma_i\, \X(0),
\,\bar\X\,\Gamma_j \left( S^\dagger h \right)_{x_-} \right\}|\bar{B}\rangle.\nonumber\\
\end{eqnarray}
The leading order Lagrangian does not contain any interactions between
hard-collinear and soft fields. Since the $B$-meson states contain only
soft particles, we should take the vacuum matrix element of the
hard-collinear fields. From equation (\ref{eq:Wij_n}) we seem to have two
such matrix elements, but these are related by the $PT$ symmetry of the
strong interaction, as explained in appendix \ref{app:A}.
Consequently, we have
\begin{eqnarray}
\label{eq:Jn_def} \frac{\nslash}{2}\delta_{kl}\,{\cal J}_{n}(p^2)&\stackrel{\rm def.}{=}&
\int\,d^4x\, e^{-ip\cdot x}\langle\Omega|\, T\left\{\X_k(0),
\, \left[\bar\X\,n\cdot\A_{hc}\right]_l(x)\right\}|\Omega\rangle\nonumber\\
&\stackrel{PT}{=}& \int\,d^4x\, e^{-ip\cdot x}\langle\Omega|\,
T\left\{\left[n\cdot\A_{hc}\,\X\right]_l(0), \,\bar\X_k(x)
\right\}|\Omega\rangle,
\end{eqnarray}
where $k,l$ are color indices. Inserting this definition into
(\ref{eq:Wij_n}), we have
\begin{eqnarray}
W_{ij}&=&-\,{\rm
Im}\,\int\,d^4x\, e^{i(q-m_bv)\cdot x}\int \frac{d^4p}{(2\pi)^4}\,e^{ip\cdot x}\,\frac{i}{\pi}\,{\cal J}_{n}(p^2)\nonumber\\
&&\frac{1}{m_b}\frac{1}{2 M_B}\langle\bar{B}|\left[\left(\bar h S\right)_0\,\Gamma_i\,\frac{\nslash}{2}
\Gamma_j\frac{\nslash}{2}\, \left( S^\dagger h \right)_{x_-}+\left(\bar h S\right)_0\,\frac{\nslash}{2}\Gamma_i \,\frac{\nslash}{2}\Gamma_j
\left( S^\dagger h \right)_{x_-} \right]|\bar{B}\rangle.
\end{eqnarray}
Defining the subleading jet function $j_n$ as
\begin{equation}
j_n(p^2)=\frac1\pi{\rm Im}\left[i\,{\cal J}_{n}(p^2)\right],
\end{equation}
and using the definition of the leading order shape function \cite{Bosch:2004cb,Bosch:2004th}
\begin{equation}
\frac{\langle\bar{B}(v)|\left(\bar h S\right)_0\,\Gamma
\left( S^\dagger h \right)_{x_-} |\bar{B}(v)\rangle}{2 M_B}=\frac12\,{\rm tr}\left(\Gamma\,\frac{1+\vslash}{2}\right)
\int \, d\omega\, e^{-\frac{i}{2}\omega\bar n\cdot x}\,S(\omega),
\end{equation}
we finally find the factorization formula
\begin{equation}
W_{ij}=-\frac1{8m_b}\,{\rm tr}\left[\,\Gamma_i\,\nslash\,
\Gamma_j\,\nslash\right]\int\,d\omega\, j_n(p_\omega^2)\,S(\omega)\,,
\end{equation}
which is of the general form of equation (\ref{eq:fact}).

An explicit one loop calculation of the bare subleading jet function
$j_n$, which can also be extracted from the sum of equations
(\ref{eq:nA}) and (\ref{eq:nAl}), gives
\begin{equation}
j_n(p^2)=\theta(p^2)\frac{C_F\alpha_s}{4\pi}\cdot 4\,\left(\frac1{\epsilon}+\frac54+\ln\frac {\mu^2}{p^2}\right).
\end{equation}

\subsubsection{\boldmath $j_{n'}$}
This jet function arises from the TOP of the leading order
current $J^{(0)} =e^{-im_b v\cdot x}\,\bar\X\,\Gamma \left(
S^\dagger h \right)_{x_-}$ and the second order current $J^{(2)}
=\mbox{}- e^{-im_b v\cdot x}\,\bar\X\,\Gamma\,\frac{1}{i\bar
n\cdot\partial}\, n\cdot\A_{hc} \left( S^\dagger h \right)_{x_-} $.
We define
\begin{eqnarray}
\label{eq:Jn'def} \frac1{\bar n\cdot
p}\,\frac{\nslash}{2}\delta_{kl}\,{\cal J}_{n'}(p^2)&=& \int\,d^4x\,
e^{-ip\cdot x}\langle\Omega|\, T\left\{\X_k(0), \,
\left[\bar\X\,\frac{1}{i\bar
n\cdot\partial}\,n\cdot\A_{hc}\right]_l(x)\right\}|\Omega\rangle\nonumber\\
&=& \int\,d^4x\, e^{-ip\cdot x}\langle\Omega|\,
T\left\{\left[n\cdot\A_{hc}\,\frac{1}{-i\bar
n\cdot\overleftarrow{\partial}}\,\X\right]_l(0), \,\bar\X_k(x)
\right\}|\Omega\rangle,
\end{eqnarray}
where $k,l$ are color indices. We define the subleading jet function
$j_{n'}$ as,
\begin{equation}
j_{n'}(p^2)=\frac1\pi{\rm Im}\left[i\,{\cal J}_{n'}(p^2)\right].
\end{equation}
The contribution of $j_{n'}$ to the hadronic tensor is
\begin{equation}
W_{ij}=-\frac1{\bar n\cdot p}\,{\rm
tr}\left[\,\Gamma_i\,\frac{\nslash}{2}\,
\Gamma_j\,\frac{1+\vslash}{2}\right]\int\,d\omega\,
j_{n'}(p_\omega^2)\,S(\omega)\,.
\end{equation}
An explicit one loop calculation of the bare subleading jet function
$j_{n'}$, which can also be extracted from the sum of equations
(\ref{eq:n'A}) and (\ref{eq:n'Al}), gives
\begin{equation}
j_{n'}(p^2)=\theta(p^2)\frac{C_F\alpha_s}{4\pi}\cdot
6\,\left(\frac1{\epsilon}-1+\ln\frac {\mu^2}{p^2}\right).
\end{equation}

\subsubsection{\boldmath $j^S_{11}$ and $j^A_{11}$}
\label{subsub:J11}
These two jet functions arise when combining two first order
currents. Recall from (\ref{eq:SCET_currents}) that the first order
current is
\begin{equation}\label{eq:J1}
J^{(1)} =\mbox{}- e^{-im_b v\cdot
x}\,\bar\X\,\frac{\nbslash}{2}\,\calAslash_{\perp hc}\,
    \frac{1}{i\bar n\cdot\overleftarrow{\partial}}\,\Gamma
    \left( S^\dagger h \right)_{x_-}
    - e^{-im_b v\cdot x}\,\bar\X\,\Gamma\,\frac{\nslash}{2m_b}\,
    \calAslash_{\perp hc} \left( S^\dagger h \right)_{x_-}\,.
\end{equation}

Since there are two terms in the first order current, one might
naively assume that we will need to define a different jet function
for each pair of terms. This is not the case, as we explain in detail
in appendix \ref{app:B}. Schematically, the reason is that the inverse
derivative acts on \emph{all} the hard-collinear fields in the first
term of (\ref{eq:J1}), which translates into an overall $1/\bar n\cdot
p$ factor.

For all the terms we need to consider the following decomposition of
the TOP of hard-collinear fields,
\begin{eqnarray}
\label{eq:J11def} && \int\,d^4x\, e^{-ip\cdot x}\langle\Omega|\,
T\Big\{\Big[\A^\mu_{\perp
hc}\,\X\Big]_k(0)\,,\Big[\bar\X\,\A^\nu_{\perp\,hc}\Big]_l(x)\Big\}|\Omega\rangle=
\nonumber\\&& \bar n\cdot
p\,\frac{\nslash}{2}\frac{g_\perp^{\mu\nu}}{d-2}\,\delta_{kl}\,{\cal
J}^S_{11}(p^2)+\bar n\cdot
p\,\frac{\nslash}{2}\,\gamma_5\frac{i\epsilon_\perp^{\mu\nu}}{d-2}\,\delta_{kl}\,{\cal
J}^A_{11}(p^2).
\end{eqnarray}
It is clear that $g_\perp^{\mu\nu}$ and $i\epsilon_\perp^{\mu\nu}$
are the only possible tensors the TOP can depend on. The Dirac
structure that accompanies each tensor is determined by $PT$
invariance (see appendix \ref{app:A}). We now define as usual,
\begin{eqnarray}
j^S_{11}(p^2)=\frac1\pi{\rm Im}\left[i\,{\cal
J}^S_{11}(p^2)\right]\quad \mbox{and}\quad
j^A_{11}(p^2)=\frac1\pi{\rm Im}\left[i\,{\cal J}^A_{11}(p^2)\right].
\end{eqnarray}

By an explicit calculation one can show that while $j^S_{11}$ starts
${\cal O} (\alpha_s)$, $j^A_{11}$ is non-zero only at ${\cal O}
(\alpha_s^2)$. The one loop result for $j^S_{11}$ is
\begin{equation}
j^S_{11}(p^2)=-\theta(p^2)\frac{C_F\alpha_s}{4\pi}.
\end{equation}
The contribution of $j^S_{11}$ and $j^A_{11}$ to the hadronic tensor is
\begin{eqnarray}
W_{ij}&=&-\frac{\bar n \cdot p}{16m_b^2}\,\mbox{tr}\left[\Gamma_i\,\nslash\,\Gamma_j\,\nslash\right]\int\,d\omega\,
\,j^S_{11}(p_\omega^2)\,S(\omega)-\frac{\bar n \cdot p}{16m_b^2}\,\mbox{tr}\left[\Gamma_i\,\nslash\gamma_5\,\Gamma_j\,\nslash\gamma_5\right]\int\,d\omega\,
\,j^A_{11}(p_\omega^2)\,S(\omega)\nonumber\\
&-&\Bigg\{\frac1{4\bar n\cdot p}\,\mbox{tr}\left[\Gamma_i\,\nbslash\,\Gamma_j\,\frac{1+\vslash}{2}\right]-
\frac{1}{16m_b}\,\mbox{tr}\left[ \Gamma_i\,\gamma_\rho^\perp\,
        \Gamma_j\,\gamma_\perp^\rho\right]-\frac{1}{16m_b}\,\mbox{tr}\left[ \Gamma_i\,\gamma_\rho^\perp\gamma_5\,
        \Gamma_j\,\gamma_\perp^\rho\gamma_5 \right]\Bigg\}\nonumber\\
&\times&\int\,d\omega\,
\,\bigg[j^S_{11}(p_\omega^2)+j^A_{11}(p_\omega^2)\bigg]\,S(\omega).
\end{eqnarray}.

\subsubsection{\boldmath $j_{G}$ and $j_{K}$}
This jet function arises from the TOP of the leading order
current $J^{(0)} =e^{-im_b v\cdot x}\,\bar\X\,\Gamma \left( S^\dagger
h \right)_{x_-}$ and the second order current $J^{(2)} =\mbox{}-
e^{-im_b v\cdot x}\,\bar\X\,\Gamma\,\frac{1}{i\bar n\cdot\partial}\,
\frac{(i\calDslash_{\perp hc}\,\calAslash_{\perp hc})}{m_b} \left(
S^\dagger h \right)_{x_-} $. We find it more useful to use the
original form of the current as it appeared in
\cite{Beneke:2002ph,Beneke:2002ni}. Using the identity
\begin{equation}
i\Dslash_{\perp}\,i\Dslash_{\perp}=\left(iD_\perp\right)^2+\frac{g}2\,\sigma^\perp_{\mu\nu}G_\perp^{\mu\nu},
\end{equation}
we find  that the current can be written as,
\begin{eqnarray}
&&\bar\X\,\Gamma\,\frac{1}{i\bar n\cdot\partial}\,
    \frac{(i\calDslash_{\perp hc}\,\calAslash_{\perp hc})}{m_b}
    \left( S^\dagger h \right)_{x_-}=\bar\xi\,\Gamma\,\frac{1}{i\bar n\cdot D}\,\frac{\big[i\Dslash_{\perp hc}\,i\Dslash_{\perp hc}\,W\big]}{m_b}\left( S^\dagger h \right)_{x_-}
 =\nonumber\\
&&\bar\X\,\Gamma\,\frac{1}{i\bar n\cdot\partial}\,
    \frac{\big[W^\dagger\,\left(iD_{\perp hc}\right)^2\,W+\frac{g}{2}
    W^\dagger\,\sigma_{\mu\nu}G^{\mu\nu}\,W\big]}{m_b}
    \left( S^\dagger h \right)_{x_-}.
\end{eqnarray}
Since $\left(iD_{\perp hc}\right)^2$ is even under $PT$ symmetry, while
$G^{\mu\nu}$ is odd, we need to separate the two parts of this
current.  We define,
\begin{eqnarray}
\label{eq:JKdef} \frac{\nslash}{2}\delta_{kl}\,{\cal J}_{K}(p^2)&=& \int\,d^4x\,
e^{-ip\cdot x}\langle\Omega|\, T\left\{\X_k(0), \,
\left[\bar\X\,\frac{1}{i\bar
n\cdot\partial}\,W^\dagger\,\left(iD_{\perp hc}\right)^2\,W\right]_l(x)\right\}|\Omega\rangle\nonumber\\
&=& \int\,d^4x\, e^{-ip\cdot x}\langle\Omega|\,
T\left\{\left[W^\dagger\,\left(i\overleftarrow{D}_{\perp hc}\right)^2\,W\,\frac{1}{-i\bar
n\cdot\overleftarrow{\partial}}\,\X\right]_l(0), \,\bar\X_k(x)
\right\}|\Omega\rangle,
\end{eqnarray}

\begin{eqnarray}
\label{eq:JGdef} \frac{i\epsilon_\perp^{\mu\nu}}{d-2}\,\frac{\nslash}{2}\gamma_5\,\delta_{kl}\,{\cal J}_{G}(p^2)&=& \int\,d^4x\,
e^{-ip\cdot x}\langle\Omega|\, T\left\{\X_k(0), \,
\left[\bar\X\,\frac{1}{i\bar
n\cdot\partial}\,W^\dagger\,G^{\mu\nu}\,W\right]_l(x)\right\}|\Omega\rangle\nonumber\\
&=& \int\,d^4x\, e^{-ip\cdot x}\langle\Omega|\,
T\left\{\left[W^\dagger\,G^{\mu\nu}\,W\,\frac{1}{-i\bar
n\cdot\overleftarrow{\partial}}\,\X\right]_l(0), \,\bar\X_k(x)
\right\}|\Omega\rangle,
\end{eqnarray}

and the corresponding subleading jet functions,

\begin{eqnarray}
j_K(p^2)=\frac1\pi{\rm Im}\left[i\,{\cal
J}_K(p^2)\right]\quad \mbox{and}\quad
j_G(p^2)=\frac1\pi{\rm Im}\left[i\,{\cal J}_{G}(p^2)\right].
\end{eqnarray}

Their contribution to the hadronic tensor is

\begin{eqnarray}
W_{ij}&=&-\frac{1}{m_b}\,{\rm
tr}\left[\,\Gamma_i\,\frac{\nslash}{2}\,
\Gamma_j\,\frac{1+\vslash}{2}\right]\int\,d\omega\,
j_{K}(p_\omega^2)\,S(\omega)\nonumber\\
&&-\frac{1}{16 m_b}\,{\rm
tr}\big[\,\Gamma_i\,\nslash\gamma_5\,
\Gamma_j\,\left(\nbslash-\nslash\right)\gamma_5\big]\int\,d\omega\,
j_{G}(p_\omega^2)\,S(\omega).
\end{eqnarray}

An explicit one loop calculation gives,

\begin{eqnarray}
j_{K}(p^2)&=&\theta(p^2)\frac{C_F\alpha_s}{4\pi}\cdot
(-2)\,\left(\frac1{\epsilon}+\frac54+\ln\frac {\mu^2}{p^2}\right)\nonumber\\
j_{G}(p^2)&=&-\theta(p^2)\frac{C_F\alpha_s}{4\pi}.
\end{eqnarray}

Notice the interesting fact that at one loop $2j_{K}+j_n=0$. Similarly
$2{\cal J}_K+{\cal J}_n=0$, which is true even without expanding in
$\epsilon$. It is unclear whether this is a one loop ``accident" or a
more general result that holds to all orders in perturbation theory.

\subsubsection{\boldmath $j_{A}$ and $j_{S}$}
This jet function arises from the TOP of the leading order
current $J^{(0)} =e^{-im_b v\cdot x}\,\bar\X\,\Gamma \left( S^\dagger
h \right)_{x_-}$ and the second order current $J^{(2)} =\mbox{}-
e^{-im_b v\cdot x}\,\bar\X\,\frac{i\overleftarrow{\calDslash\!}_{\perp hc}}{m_b}\,
    \frac{1}{i\bar n\cdot\overleftarrow{\partial}}\,\frac{\nbslash}{2}\,
    \Gamma\,\frac{\nslash}{2}\,\calAslash_{\perp hc}
    \left( S^\dagger h \right)_{x_-}$. We define
\begin{eqnarray}
&& \int\,d^4x\, e^{-ip\cdot x}\langle\Omega|\,
T\Big\{\X_k(0),\Big[\bar\X\,i\overleftarrow{\cal D}^\mu_{\perp hc}\frac{1}{i\bar n\cdot\overleftarrow{\partial}}\,\A^\nu_{\perp\,hc}\Big]_l(x)\Big\}|\Omega\rangle=
\nonumber\\
&& \frac{\nslash}{2}\frac{g_\perp^{\mu\nu}}{d-2}\,\delta_{kl}\,{\cal
J}_{S}(p^2)+\frac{\nslash}{2}\,\gamma_5\frac{i\epsilon_\perp^{\mu\nu}}{d-2}\,\delta_{kl}\,{\cal
J}_A(p^2).
\end{eqnarray}
$PT$ symmetry implies that
\begin{eqnarray}
\label{eq:Jlast}
&& \int\,d^4x\, e^{-ip\cdot x}\langle\Omega|\,
T\Big\{\Big[\A^\nu_{\perp\,hc}\,
\frac{1}{i\bar n\cdot\partial}\,i{\cal D}^\mu_{\perp hc}\,\X\Big]_l(0),\bar{\X}_k(x)\Big\}|\Omega\rangle=
\nonumber\\
&& \frac{\nslash}{2}\frac{g_\perp^{\mu\nu}}{d-2}\,\delta_{kl}\,{\cal
J}_{S}(p^2)-\frac{\nslash}{2}\,\gamma_5\frac{i\epsilon_\perp^{\mu\nu}}{d-2}\,\delta_{kl}\,{\cal
J}_A(p^2).
\end{eqnarray}
Notice the minus sign in front of the second term.

We now define the corresponding subleading jet functions,
\begin{eqnarray}
j_S(p^2)=\frac1\pi{\rm Im}\left[i\,{\cal
J}_S(p^2)\right]\quad \mbox{and}\quad
j_A(p^2)=\frac1\pi{\rm Im}\left[i\,{\cal J}_A(p^2)\right].
\end{eqnarray}

Their contribution to the hadronic tensor is
\begin{eqnarray}
W_{ij}&=&\frac{1}{16m_b}\Bigg\{\mbox{tr}\left[ \Gamma_i\,\gamma_\rho^\perp\,
        \Gamma_j\,\gamma_\perp^\rho\right] -
    \mbox{tr}\left[ \Gamma_i\,\gamma_\rho^\perp\gamma_5\,
        \Gamma_j\,\gamma_\perp^\rho\gamma_5 \right] \Bigg\}\nonumber\\
&\times&
\int\,d\omega\,\left[\,j_S(p_\omega^2)+(3-d)\,j_A(p_\omega^2)\right]\,S(\omega).
\end{eqnarray}

An explicit one loop calculation gives,
\begin{eqnarray}
j_{S}(p^2)&=&\theta(p^2)\frac{C_F\alpha_s}{4\pi}\cdot
\left(-\frac32\right)\nonumber\\
j_{A}(p^2)&=&\theta(p^2)\frac{C_F\alpha_s}{4\pi}\cdot\left(\frac12\right).
\end{eqnarray}

\subsection{Renormalization}
\label{subsec:ren}
Having defined and calculated all the subleading jet function to one
loop order, we are ready to discuss their renormalization. We note
first that only 3 out of the 8 functions we have defined require
renormalization at one loop order. These are
\begin{eqnarray}
j^{\rm bare}_n(p^2,\mu)&=&\theta(p^2)\frac{C_F\alpha_s}{4\pi}\cdot 4\,\left(\frac1{\epsilon}+\frac54+\ln\frac {\mu^2}{p^2}\right)+{\cal O}(\alpha_s^2)\nonumber\\
j^{\rm bare}_{n'}(p^2,\mu)&=&\theta(p^2)\frac{C_F\alpha_s}{4\pi}\cdot 6\,\left(\frac1{\epsilon}-1+\ln\frac {\mu^2}{p^2}\right)+{\cal O}(\alpha_s^2)\nonumber\\
j^{\rm bare}_K(p^2,\mu)&=&\theta(p^2)\frac{C_F\alpha_s}{4\pi}\cdot
\big(-2\big)\,\left(\frac1{\epsilon}+\frac54+\ln\frac
{\mu^2}{p^2}\right)+{\cal O}(\alpha_s^2).
\end{eqnarray}
In order to renormalize these function we have to introduce a \emph{new} function
\begin{equation}
j_0(p^2)=\theta(p^2)+{\cal O}(\alpha_s),
\end{equation}
where only the \emph{zeroth} order in $\alpha_s$ part of $j_0(p^2)$
is needed in order to renormalize the subleading jet functions at
\emph{first} order in $\alpha_s$.

 It is very tempting to relate this function to the integral
over the leading order jet function, namely to identify $j_0(p^2)$
with
\begin{equation}
\label{eq:intJ}
\int^{p^2}\,dp'^{\,2}\,J(p'^{\,2},\mu)
\end{equation}
(where the lower limit of the integral can be any negative number),
since both are equal at ${\cal O}(\alpha_s^0)$. A very similar function, $j(\ln \frac{p^2}{\mu^2}, \mu)$ was defined in \cite{Becher:2006qw}
\begin{equation}
j(\ln \frac{p^2}{\mu^2},\mu)=\int_0^{p^2}\,dp'^{\,2} \,J(p'^{\,2},\mu).
\end{equation}
In that paper the authors derived the two loop expression for $j$ and
its anomalous dimension. Since we only use the ${\cal O}(\alpha_s^0)$
expression for $j_0$, we will refrain from identifying it with
equation (\ref{eq:intJ}) or with $j$ of \cite{Becher:2006qw}.  It
seems plausible that all these expression are the same, but in order
to determine whether they coincide beyond ${\cal O}(\alpha_s^0)$, a
two loop calculation of the subleading jet function is needed. Such a
calculation is beyond the scope of this paper.

The most convenient scheme in which to renormalize the subleading
jet functions (as well as the subleading shape functions) is the
$\overline{\rm DR}$ subtraction scheme
\cite{Siegel:1979wq,Capper:1979ns}. In this scheme the Dirac algebra
is performed in $d=4$ dimensions, while loop integrals are evaluated
in $d=4-2\epsilon$ dimensions. Choosing this scheme ensures that the
renormalized subleading functions are the same for $\bar B \to X_u\,
l\,\bar\nu$ and the $Q_{7\gamma}-Q_{7\gamma}$ contribution to $\bar B \to
X_s\,\gamma$.

We renormalize the subleading jet function in the following way.
Define the matrix $Z(p^2,p'^{\,2},\mu)$ via
\begin{equation}
\left(\begin{array}{c}j_n(p^2)\\j_{n'}(p^2)\\j_K(p^2)\end{array}\right)
=\int\,dp'^{\,2}\,Z(p^2,p'^{\,2},\mu)
\left(\begin{array}{c}j^{\rm bare}_0(p'^{\,2})\\j^{\rm bare}_n(p'^{\,2})\\j^{\rm bare}_{n'}(p'^{\,2})\\j^{\rm bare}_K(p'^{\,2})\end{array}\right).
\end{equation}
At one loop order we find that in the $\overline{\rm DR}$ scheme
\begin{equation}
Z(p^2,p'^{\,2},\mu)=\delta(p^2-p'^{\,2})\left(\begin{array}{rrrr}-4a&1&0&0\\-6a&0&1&0\\2a&0&0&1\end{array}\right),
\end{equation}
where $a=C_F\alpha_s/4\pi\epsilon$. Notice that $Z(p^2,p'^{\,2},\mu)$
is not a square matrix, since we do not renormalize $j_0$. If indeed
$j_0$ is related to the integral over the leading order jet function,
then $Z$ will have to include other distributions apart from a delta
function, for details, see \cite{Becher:2006qw}.

From $Z(p^2,p'^{\,2},\mu)$ we can find the renormalization group
equations for the subleading jet functions. We have at one loop
\begin{equation}
\frac{d}{d\ln \mu}\left(\begin{array}{c}j_n(p^2)\\j_{n'}(p^2)\\j_K(p^2)\end{array}\right)
=\int\,dp'^{\,2}\,\delta(p^2-p'^{\,2})\,\frac{C_F\alpha_s}{4\pi}\,\left(\begin{array}{rrrr}8&0&0&0\\12&0&0&0\\-4&0&0&0\end{array}\right)
\left(\begin{array}{c}j_0(p'^{\,2})\\j_n(p'^{\,2})\\j_{n'}(p'^{\,2})\\j_K(p'^{\,2})\end{array}\right),
\end{equation}
The expression for the renormalized subleading jet functions are
\begin{eqnarray}
j_n(p^2,\mu)&=&\theta(p^2)\frac{C_F\alpha_s}{4\pi}\cdot 4\,\left(\frac54+\ln\frac {\mu^2}{p^2}\right)+{\cal O}(\alpha_s^2)\nonumber\\
j_{n'}(p^2,\mu)&=&\theta(p^2)\frac{C_F\alpha_s}{4\pi}\cdot 6\,\left(-1+\ln\frac {\mu^2}{p^2}\right)+{\cal O}(\alpha_s^2)\nonumber\\
j_K(p^2,\mu)&=&\theta(p^2)\frac{C_F\alpha_s}{4\pi}\cdot
\big(-2\big)\,\left(\frac54+\ln\frac
{\mu^2}{p^2}\right)+{\cal O}(\alpha_s^2).
\end{eqnarray}
It is unclear whether with the inclusion of $j_0$ the list of the
subleading jet functions closes under renormalization. In particular
one might expect that we need to define a more general subleading jet
function that depends on more than one variable. For example the
discontinuity of the Fourier transform of
\begin{equation}
\langle\Omega|\, T\left\{\X_k(0),\,n\cdot\A^{hc}_{mn}(y),\,\bar\X_l(x)\right\}|\Omega\rangle
\end{equation}
where $k,l,m,n$ are color indices. An interesting fact about this type
of subleading jet function is that at order $g^2$ it only becomes
singular in the $y\to 0$ and $y\to x$ limits, where it can be related
to $j_n$.

Clearly this topic deserves further study, but for our purposes the
renormalization via the introduction of $j_0$ is sufficient at the
order in which we are working, namely, $\alpha_s(\mu_i)/m_b$.

For phenomenological applications it is convenient to set the scale of
the subleading jet functions to be the same as the scale of the
leading order jet function. This is also the convenient scale in which
to extract the leading order shape function \cite{Lange:2005yw}. At
any case the resummation of the subleading logs is expected to be a
small effect. As a result, the renormalization group equations of the
subleading jet function are not expected to be important for
phenomenological applications.

\subsection{Results}
\label{subsec:results}
We are ready to summarize our results. The subleading jet functions'
contribution to the hadronic tensor can be written as
\begin{eqnarray}
\label{eq:SJF}
W_{ij}^{\rm \,SJF}=
-\int d\omega &\bigg[&j_{n}(p_\omega^2,\mu)\frac{2\tilde{T}_2}{m_b}\,+j_{n'}(p_\omega^2,\mu)\frac{\tilde{T}_1}{\bar n\cdot p}+j_{K}(p_\omega^2,\mu)\frac{\tilde{T}_1}{m_b}\,+j_G(p_\omega^2,\mu)\frac{\tilde{T}_4}{m_b}\nonumber\\
&+&j^S_{11}(p_\omega^2,\mu)\left(\frac{\bar n\cdot p}{m_b^2}\,\tilde{T}_2+\frac{\tilde{T}_3}{\bar n\cdot p}-\frac{\tilde{T}_5}{m_b}-\frac{\tilde{T}_6}{m_b}\right)\nonumber\\
&+&j^A_{11}(p_\omega^2,\mu)\left(\frac{\bar n\cdot p}{m_b^2}\,\tilde{T}_7+\frac{\tilde{T}_3}{\bar n\cdot p}-\frac{\tilde{T}_5}{m_b}-\frac{\tilde{T}_6}{m_b}\right)\nonumber\\
&+&j_S(p_\omega^2,\mu)\left(\frac{\tilde{T}_5}{m_b}-\frac{\tilde{T}_6}{m_b}\right)+j_A(p_\omega^2,\mu)\left(-\frac{\tilde{T}_5}{m_b}+\frac{\tilde{T}_6}{m_b}\right)
\,\bigg] S(\omega),
\end{eqnarray}
where the traces $\tilde{T}_1...\tilde{T}_6$ are\footnote{Under the
  assumption that $\Gamma_i$ and $\Gamma_j$ contain the same number of
  Dirac's gamma matrices, we can relate $\tilde{T}_1...\tilde{T}_6$ to
  $T_1...T_4$ in equation (\ref{eq:traces}). The relations are
  $T_1=\tilde{T}_1/2,\,T_2=\tilde{T}_1/2-2\tilde{T}_2,\,T_3=2\tilde{T}_6,\,T_4=\tilde{T}_4$.}
\begin{eqnarray}
\label{traces}
    \tilde{T}_1 &=& \frac12\,\mbox{tr}\left[ \Gamma_i\,\nslash\,\Gamma_j\,
            \frac{1+\vslash}{2} \right], \quad\quad
    \tilde{T}_2 = \frac{1}{16}\,\mbox{tr}\,\Big[ \Gamma_i\,\nslash\,\Gamma_j\,
            \nslash \Big],\nonumber\\
    \tilde{T}_3 &=& \frac14\,\mbox{tr}\left[ \Gamma_i\,\nbslash\,\Gamma_j\,
        \frac{1+\vslash}{2} \right],\quad\quad
    \tilde{T}_4 = \frac{1}{16}\,\mbox{tr}\left[ \Gamma_i\,\nslash\gamma_5\,\Gamma_j\,
        (\nbslash-\nslash)\,\gamma_5 \right],\nonumber\\
    \tilde{T}_5 &=& \frac{1}{16}\,\mbox{tr}\left[ \Gamma_i\,\gamma_\rho^\perp\,
        \Gamma_j\,\gamma_\perp^\rho\right] ,\quad\quad\hspace{0.2cm}
    \tilde{T}_6 = \frac{1}{16}\,\mbox{tr}\left[ \Gamma_i\,\gamma_\rho^\perp\gamma_5\,
        \Gamma_j\,\gamma_\perp^\rho\gamma_5 \right] \nonumber\\
        \tilde{T}_7 &=& \frac{1}{16}\,\mbox{tr}\,\Big[ \Gamma_i\,\nslash\gamma_5\,\Gamma_j\,
            \nslash\gamma_5 \Big].
\end{eqnarray}
The subleading jet functions $j^S_{11},j^A_{11},j_{n},j_{n'}, j_{K},
j_{G}, j_{S}, j_{A}$ are defined in section \ref{subsec:sjfdef}. The
renormalized one loop expressions for them are
\begin{eqnarray}
j^S_{11}(p^2,\mu)&=&\theta(p^2)\frac{C_F\alpha_s(\mu)}{4\pi}\,\big(-1\big)+{\cal O}(\alpha_s^2)\nonumber\\
j^A_{11}(p^2,\mu)&=&0+{\cal O}(\alpha_s^2)\nonumber\\
j_n(p^2,\mu)&=&\theta(p^2)\frac{C_F\alpha_s(\mu)}{4\pi}\,\left(5+4\ln\frac {\mu^2}{p^2}\right)+{\cal O}(\alpha_s^2)\nonumber\\
j_{n'}(p^2,\mu)&=&\theta(p^2)\frac{C_F\alpha_s(\mu)}{4\pi}\,\left(-6+6\ln\frac {\mu^2}{p^2}\right)+{\cal O}(\alpha_s^2)\nonumber\\
j_K(p^2,\mu)&=&\theta(p^2)\frac{C_F\alpha_s(\mu)}{4\pi}\,\left(-\frac52-2\ln\frac {\mu^2}{p^2}\right)+{\cal O}(\alpha_s^2)\nonumber\\
j_G(p^2,\mu)&=&\theta(p^2)\frac{C_F\alpha_s(\mu)}{4\pi}\,\big(-1\big)+{\cal O}(\alpha_s^2)\nonumber\\
j_S(p^2,\mu)&=&\theta(p^2)\frac{C_F\alpha_s(\mu)}{4\pi}\,\left(-\frac32\right)+{\cal O}(\alpha_s^2)\nonumber\\
j_A(p^2,\mu)&=&\theta(p^2)\frac{C_F\alpha_s(\mu)}{4\pi}\,\left(\frac12\right)+{\cal O}(\alpha_s^2).
\end{eqnarray}
These expressions are in the $\overline{\rm DR}$ subtraction scheme
\cite{Siegel:1979wq,Capper:1979ns}. As explained in section
\ref{subsec:ren} this is the most appropriate renormalization scheme
for the subleading jet and shape functions.

The scale dependence in equation (\ref{eq:SJF}) cancels against the
scale dependence of the subleading shape functions' contribution
\cite{Bosch:2004cb}
\begin{eqnarray}
   W_{ij}^{\rm \,SSF} &=& \int d\omega\,\delta(n\cdot p+\omega) \left[
    \frac{\omega\,S(\omega,\mu) + t(\omega,\mu)}{m_b}\,T_2
    + \frac{s(\omega,\mu)}{m_b}\,T_1
    + \frac{t(\omega,\mu)}{\bar n\cdot p}\,T_3
    + \frac{u(\omega,\mu)}{\bar n\cdot p}\,T_1\right.\nonumber\\
    &-& \left.\frac{v(\omega,\mu)}{\bar n\cdot p}\,T_4
-\pi\alpha_s\left(
    \frac{f_u(\omega,\mu)}{\bar n\cdot p}\,T_1
   + \frac{f_v(\omega,\mu)}{\bar n\cdot p}\,T_4\right)\right]+{\cal O}(\alpha_s).
\end{eqnarray}
For the definition of the subleading shape functions and the traces
$T_1...T_4$, see section \ref{sec:review}.

We now specialize to the cases of semileptonic and radiative $B$
decays, using the expressions for $\Gamma_i$ and $\Gamma_j$ given in
section \ref{sec:review}. For the $Q_{7\gamma}-Q_{7\gamma}$
contribution to $\bar B \to X_s\,\gamma$ we need the ``trace" of the
hadronic tensor: $W$. Its relation to the photon spectrum of $\bar B \to
X_s\,\gamma$ is given in equation (\ref{eq:Wiispec}). The subleading
jet functions' contribution to $W$ is
\begin{eqnarray}
W^{\rm \,SJF}=\int d\omega&\bigg[&\frac{1}{m_b}\bigg(4j_{K}(p_\omega^2,\mu)+\,4j_{n}(p_\omega^2,\mu)+2j_G(p_\omega^2,\mu)\bigg)+\frac{4}{\bar n\cdot p}\,j_{n'}(p_\omega^2,\mu)\nonumber\\
&+&\frac{2\bar n\cdot p}{m_b^2}\,\bigg(j^S_{11}(p_\omega^2,\mu)-j^A_{11}(p_\omega^2,\mu)\bigg)\bigg]\, S(\omega).
\end{eqnarray}
Note that $j_S$ and $j_A$ do not contribute to $W$.
At the lowest order in $\alpha_s$ this expression is
\begin{equation}
W^{\rm \,SJF}=\int d\omega\, \frac{1}{m_b}\,\frac{C_F\alpha_s(\mu)}{4\pi}\,\theta(p_\omega^2)\left[32\,\ln\frac {\mu^2}{p_\omega^2}-18\right] S(\omega) +{\cal O}(\alpha_s^2),
\end{equation}
where in order to simplify the expression, we have used the fact that
for this decay mode $\bar n\cdot p=m_b$. For completeness we list also
the subleading shape functions' contribution
\begin{eqnarray}
\label{eq:Wii_new}
   W^{\rm \,SSF} = -\frac2{m_b}\int d\omega\,\delta(n\cdot p+\omega)\, \Big[
    &-&\,\omega\,S(\omega,\mu) + s(\omega,\mu)-t(\omega,\mu)+u(\omega,\mu)-v(\omega,\mu)\nonumber\\
&-&\pi\alpha_s\,f_u(\omega,\mu)-\pi\alpha_s\,f_v(\omega,\mu)\Big]+{\cal O}(\alpha_s)
\end{eqnarray}

For $\bar B \to X_u\, l\,\bar\nu$ we need the three ``form factors":
$\tilde{W}_1, \tilde{W}_2$ and $\tilde{W}_{\rm
  comb}\equiv\frac{y}{4}\tilde{W}_3+\tilde{W}_4+\frac1{y}\tilde{W}_5$.
$\tilde{W}_i$ are defined in equation (\ref{eq:Widef}) and their
relation to the triple differential decay rate is given by equation
(\ref{eq:dgu}). The subleading jet functions' contribution is
\begin{eqnarray}
\tilde{W}_1^{\rm \,SJF}&=&
-\int d\omega \bigg[\frac{1}{\bar n\cdot p}\bigg(2j_{n'}(p_\omega^2,\mu)+j^S_{11}(p_\omega^2,\mu)+j^A_{11}(p_\omega^2,\mu)\bigg)\nonumber\\
&&\hspace{1.5cm}+\frac{1}{m_b}\bigg(2j_{K}(p_\omega^2,\mu)+j_G(p_\omega^2,\mu)\bigg)\bigg]\, S(\omega)\nonumber\\
\tilde{W}_2^{\rm \,SJF}&=&-\int d\omega\,\frac{2}{\bar n\cdot p} \bigg[j^S_{11}(p_\omega^2,\mu)+j^A_{11}(p_\omega^2,\mu)\bigg]\, S(\omega)\nonumber\\
\tilde{W}_{\rm comb}^{\rm SJF}&=&-\int d\omega \bigg[\left(\frac{4}{m_b}-\frac{2}{\bar n\cdot p}\right)\bigg(j^S_{11}(p_\omega^2,\mu)+j^A_{11}(p_\omega^2,\mu)\bigg)\nonumber\\
&&\hspace{1.5cm}+\frac{2}{\bar n\cdot p}\bigg(j_{n}(p_\omega^2,\mu)-j_{G}(p_\omega^2,\mu)\bigg)\bigg]\, S(\omega).
\end{eqnarray}
At the order in which we are working in, it can approximated by $\bar n\cdot p/m_b$. As for $W$, $j_S$ and $j_A$ do not contribute to $\tilde{W}_1,\,\tilde{W}_2$ or $\tilde{W}_{\rm comb}$.
At the lowest order in $\alpha_s$ the last equation is
\begin{eqnarray}
\label{eq:Wi_new}
\tilde{W}_1^{\rm \,SJF}&=&
-\int d\omega \frac{C_F\alpha_s(\mu)}{4\pi}\,\theta(p_\omega^2) \bigg[\frac{1}{\bar n\cdot p}\left(12\ln\frac {\mu^2}{p_\omega^2}-11\right)-
\frac{1}{m_b}\left(4\ln\frac {\mu^2}{p_\omega^2}+6\right)\bigg]\, S(\omega)\nonumber\\
\tilde{W}_2^{\rm \,SJF}&=&\int d\omega\, \frac{C_F\alpha_s(\mu)}{4\pi}\,\theta(p_\omega^2)\,\frac{2}{\bar n\cdot p}\, S(\omega)\nonumber\\
\tilde{W}_{\rm comb}^{\rm SJF}&=&-\int d\omega  \frac{C_F\alpha_s(\mu)}{4\pi}\,\theta(p_\omega^2)\bigg[
\frac{1}{\bar n\cdot p}\left(8\ln\frac {\mu^2}{p_\omega^2}+14\right)-\frac{4}{m_b}\bigg]\, S(\omega).
\end{eqnarray}
If we are using the so called ``BLNP" approach \cite{Lange:2005yw},
i.e. using the definition of $y$ as in equation (\ref{eq:ydef}), then
some of the subleading terms are absorbed into the leading order
formula (\ref{eq:BuLP}). The subleading jet functions' contribution to
$\tilde{W}_1$ in this case is given by
\begin{equation}
\label{eq:W1_BLNP}
\tilde{W}^{\rm \,SJF}_{1,\,\rm BLNP}=
-\int d\omega \frac{C_F\alpha_s(\mu)}{4\pi}\,\theta(p_\omega^2) \bigg[\frac{1}{\bar n\cdot p}\left(12\ln\frac {\mu^2}{p_\omega^2}-15\right)-
\frac{1}{m_b}\left(4\ln\frac {\mu^2}{p_\omega^2}+2\right)\bigg]\, S(\omega),
\end{equation}
where there is no change to $\tilde{W}_2$ and $\tilde{W}_{\rm comb}$. For completeness we list also the subleading shape functions' contribution
\begin{eqnarray}
\tilde{W}_1^{\rm \,SSF}&=&\int d\omega\,\delta(n\cdot p+\omega)\left[\frac{\omega\,S(\omega,\mu)+s(\omega,\mu)+t(\omega,\mu)}{m_b}+\frac{u(\omega,\mu)-v(\omega,
\mu)}{\bar n\cdot p}\right]\nonumber\\
\tilde{W}_2^{\rm \,SSF}&=&0\nonumber\\
\tilde{W}_{\rm comb}^{\rm \,SSF}&=&-2\int d\omega\,\delta(n\cdot p+\omega)\left[\frac{\omega\,S(\omega,\mu)+2t
(\omega,\mu)}{\bar n\cdot p}-\frac{t(\omega,\mu)+v(\omega,\mu)}{y\,\bar n\cdot p}\right].
\end{eqnarray}
\section{Summary and Conclusions}
\label{sec:conclusions}
Decay rates of inclusive $B$ decays, namely $\bar B \to X_u\, l\,\bar\nu$
and the $Q_{7\gamma}-Q_{7\gamma}$ contribution to $\bar B \to X_s\,\gamma$,
are known to factorize in the end point region at the leading order in
$\Lambda_{\rm QCD}/m_b$ into a product of a hard function and a
universal leading order jet function convoluted with a universal
leading order shape function. The hard and jet function are calculable
in perturbation theory while the shape function is a non perturbative
object. Recently the hard function for semileptonic decays was
calculated at ${\cal O}(\alpha_s^2)$ \cite{Bonciani:2008wf,
  Asatrian:2008uk, Beneke:2008ei, Bell:2008ws}. Together with the
already known two loop calculation of the jet function
\cite{Becher:2006qw}, this will allow us to reach a full ${\cal
  O}(\alpha^2_s$) accuracy, at leading power, in describing the decay
rates in general, and in extracting $|V_{ub}|$ in particular.

Beyond leading order in $\Lambda_{\rm QCD}/m_b$, one would expect the
decay rate to factorize into sums of products of subleading hard
functions and subleading jet functions convoluted with subleading
shape functions. Of these power suppressed corrections, only the
subleading shape functions were known. In this paper we have analyzed
the subleading jet functions' contribution. These arise first at order
${\cal O}(\alpha_s)$ and appear in the partial rate convoluted with
the leading order shape function.

First, we have argued that at order $\Lambda_{\rm QCD}/m_b$ only
subleading jet and shape functions contribute. Subleading hard
function can only appear when multiplied with subleading jet or shape
functions. We have demonstrated this explicitly at one loop by showing
that the ${\cal O}(\alpha_s)$ corrections which are $\Lambda_{\rm
  QCD}/m_b$ suppressed in the end point region, arise from two
momentum regions: a hard-collinear region and a soft region. We have
then shown that the soft region is accounted for in the parton model
by the one loop matrix elements of the \emph{known} ``tree level"
subleading shape functions. The hard-collinear region is accounted for
in the parton model by the time ordered products of subleading SCET
currents that apart from the heavy quark itself, do not include any
soft fields or soft covariant derivatives.

In the main section of this paper, section \ref{sec:subjet}, we have
defined to all orders in $\alpha_s(\mu_i)$, for the case of a tree
level hard function, the 8 subleading jet functions that can
contribute to partial rates of inclusive $B$ decays, and calculated
their one loop expressions. After a short discussion of the
renormalization of the subleading jet functions, we specialized to the
phenomenologically interesting cases of $\bar B \to X_u\, l\,\bar\nu$ and
the $Q_{7\gamma}-Q_{7\gamma}$ contribution to $\bar B \to X_s\,\gamma$ and
presented explicit expressions for these decay modes. The main results
of the paper are collected in section \ref{subsec:results}.

We can now summarize the factorization formula for inclusive $B$
decays in the following way
$$
d\Gamma\sim\overbrace{H\cdot J\otimes S+\frac{1}{m_b}\sum_{i}\,h\cdot J_0\otimes s_i}^{\rm known}
+\overbrace{\frac{1}{m_b}\sum_{k}\,h\cdot j_k\otimes S}^{\rm new}
+\,{\cal O}\left(\frac{1}{m_b^2}\right),
$$
where the label ``new" refers to the new results of this paper. The
label ``known" refers to terms in the factorization formula for which
we have \emph{explicit} expressions for all of their perturbative
components. Thus $H$ is the leading order hard function, $J$ is the
leading order jet function, both known at ${\cal O}(\alpha_s^2)$,
$J_0$ is the ${\cal O}(\alpha_s^0)$ part of the leading order jet
function, $h=1+{\cal O}(\alpha_s)$, and $j_k$ are given in section
\ref{subsec:results}.  The rest of the $1/m_b$ suppressed terms for
which we do not have such explicit expressions can be found in
\cite{Lee:2004ja} and \cite{Beneke:2004in}.

While the subleading jet functions' contribution is both $\alpha_s$
and $1/m_b$ suppressed in the end point region, their contribution
becomes more important as one moves out of the this region, since the
$1/m_b$ suppression is reduced as one is integrating over larger and
larger portions of phase space. The \emph {one-loop} subleading shape
functions contribution, although formally $\alpha_s/m_b$ suppressed in
the end point region, is expected to become even more power suppressed
outside of the end point region. Furthermore, the kinematical area
outside of the end point region is becoming more important with the
constant improvement of experiments and the relaxation of kinematical
cuts. Together with the recent calculation of the leading order hard
function for semileptonic decays at ${\cal O}(\alpha^2_s)$, this work
takes another step towards a more precise description of inclusive $B$
decays in the end point region.

Although we have not presented any kind of numerical analysis, the
implementation of the subleading jet function analysis with the
framework of ``BLNP" \cite{Lange:2005yw} is relatively easy.  One
needs to replace equation (\ref{eq:Wii}) by (\ref{eq:Wii_new}) and
(\ref{eq:kinNLO}) by equations (\ref{eq:Wi_new}) and
(\ref{eq:W1_BLNP}). At the same time one needs to modify the treatment
of the subleading shape function in \cite{Lange:2005yw}. Only the
combined numerical analysis would be meaningful. Such a study is left
for future work.

Another issue that deserves further consideration is the
renormalization of the subleading jet functions. In particular, it
would be desirable to have a complete basis of subleading jet
functions. As in the case of the subleading shape functions
\cite{Trott:2005vw}, it seems that further study is needed if we wish
to understand the renormalization and mixing of these non local
operators.

Finally, subleading jet functions arise also outside of flavor
physics, for example, in the $x\to 1$ region of deep inelastic
scattering. More specifically, the subleading jet function $j_{11}^S$
appear in the factorization formula for the longitudinal structure
function \cite{Akhoury:1998gs, Akhoury:2003fw, Chay:2005rz,
  Becher:2006mr}\footnote{Note though that the subleading jet function
  in \cite{Akhoury:1998gs, Akhoury:2003fw} and \cite{Chay:2005rz}
  differ from our definition and from each other.}. A more detailed
study of the subleading jet functions' contributions to the $x\to 1$
region of deep inelastic scattering is left for future work.

\vspace{0.3cm}

{\em Acknowledgments:} I would like to thank Matthias Neubert for
suggesting the decomposition (\ref{eq:lnr}) and for useful
discussions. I would also like to thank Thomas Becher, Bj\"orn Lange,
Donal O'Connell, and Ben Pecjak for useful discussions and their
comments on the manuscript. This work is supported in part by the
Department of Energy grant DE-FG02-90ER40542 and by the United
States-Israel Bi-national Science Foundation grant 2006280.

\begin{appendix}
\section{Consequences of $PT$ symmetry}
\label{app:A}
In this section we will prove our claim that $PT$ symmetry and translation invariance allows us to relate the TOP of two different hard-collinear operators. Define
\begin{eqnarray}
T_{ab}&=&\int d^4x\, e^{-ipx}\, \langle \Omega |T\left\{O_a^\dagger(0),O_b(x) \right\}|\Omega\rangle=\nonumber\\
&=&\int d^4x\, e^{-ipx}\Big[\theta(x^0)\langle \Omega|O_b(x)O_a^\dagger(0)|\Omega\rangle\pm\theta(-x^0)\langle \Omega|O_a^\dagger(0)O_b(x)|\Omega\rangle\Big]\nonumber\\
T_{ba}&=&\int d^4x\, e^{-ipx}\, \langle \Omega |T\left\{O_b^\dagger(0),O_a(x) \right\}|\Omega\rangle=\nonumber\\
&=&\int d^4x\, e^{-ipx}\Big[\theta(x^0)\langle \Omega|O_a(x)O_b^\dagger(0)|\Omega\rangle\pm\theta(-x^0)\langle \Omega|O_b^\dagger(0)O_a(x)|\Omega\rangle\Big]
\end{eqnarray}
We would like to relate $T_{ab}$ and $T_{ba}$. Using translation invariance, and the $PT$ invariance of the strong interactions and of the vacuum, we have
\begin{eqnarray}
\langle \Omega|O_a(x)O_b^\dagger(0)|\Omega\rangle&=&\langle \Omega|O_a(0)O_b^\dagger(-x)|\Omega\rangle=\langle \Omega|\bigg[\Big(O_b^{\dagger}\Big)^{PT}\bigg]^\dagger(x)\bigg[\Big(O_a\Big)^{PT}\bigg]^{\dagger}(0)|\Omega\rangle\nonumber\\
\langle \Omega|O^\dagger_b(0)O_a(x)|\Omega\rangle&=&\langle \Omega|O^\dagger_b(-x)O_a(0)|\Omega\rangle=\langle \Omega|\bigg[\Big(O_a\Big)^{PT}\bigg]^\dagger(0)\bigg[\Big(O_b^\dagger\Big)^{PT}\bigg]^{\dagger}(x)|\Omega\rangle.
\end{eqnarray}
In order to relate $T_{ab}$ and $T_{ba}$ we need to relate
\begin{eqnarray}
\langle \Omega|\bigg[\Big(O_b^{\dagger}\Big)^{PT}\bigg]^\dagger(x)\bigg[\Big(O_a\Big)^{PT}\bigg]^{\dagger}(0)|\Omega\rangle \quad&{\rm to}&\quad\langle \Omega|O_b(x)O_a^\dagger(0)|\Omega\rangle\quad \nonumber\\
\langle \Omega|\bigg[\Big(O_a\Big)^{PT}\bigg]^\dagger(0)\bigg[\Big(O_b^\dagger\Big)^{PT}\bigg]^{\dagger}(x)|\Omega\rangle\quad&{\rm to}&\quad \langle \Omega|O_a^\dagger(0)O_b(x)|\Omega\rangle.
\end{eqnarray}
As our first example consider ${\cal J}_{n}$. In section \ref{sec:subjet} we defined,
\begin{eqnarray}
\label{eq:Jn_repeat}
&&\left(\frac{\nslash}{2}\right)_{ab}\,\delta_{kl}\,{\cal J}_{n}(p^2)=\nonumber\\
&=&\int\,d^4x\, e^{-ip\cdot x}\,\Big[\theta(x^0)\langle\Omega|\left[\bar\X\,n\cdot\A_{hc}\right]_{b\,l}(x)\X_{a\,k}(0)
|\Omega\rangle-\theta(-x^0)\langle\Omega|\X_{a\,k}(0)\left[\bar\X\,n\cdot\A_{hc}\right]_{b\,l}(x)|\Omega\rangle\Big],\nonumber\\
\end{eqnarray}
where $k,l$ are color indices and $a,b$ are spinor indices. Consider now the other TOP:
\begin{eqnarray}
&&T_{\rm other}=\int\,d^4x\, e^{-ip\cdot x}\langle\Omega|\,
T\left\{\left[n\cdot\A_{hc}\,\X\right]_{a\,l}(0), \,\bar\X_{b\,k}(x)\right\}|\Omega\rangle=\nonumber\\
&=&\int\,d^4x\, e^{-ip\cdot x}\,\Big[\theta(x^0)\langle\Omega|\bar\X_{b\,k}(x)\left[n\cdot\A_{hc}\,\X\right]_{a\,l}(0)|\Omega\rangle
-\theta(-x^0)\langle\Omega|\left[n\cdot\A_{hc}\,\X\right]_{a\,l}(0)\bar\X_{b\,k}(x)
|\Omega\rangle\Big].\nonumber\\
\end{eqnarray}
Using translation invariance and the $PT$ symmetry we have, in the
Weyl representation of Dirac $\gamma$ matrices,
\begin{eqnarray}
\label{eq:Tother_pt}
T_{\rm other}&=&\int\,d^4x\, e^{-ip\cdot x}\,\Big[\hspace{1em}\theta(x^0)\langle\Omega|\left[\bar\X\,\gamma^1\,\gamma^3\,n\cdot\A_{hc}\,\right]_{a\,l}(x)
\left[\gamma^3\,\gamma^1\,\X\right]_{b\,k}(0)|\Omega\rangle\nonumber\\
&&\hspace{6em}-\theta(-x^0)\langle\Omega|\left[\gamma^3\,\gamma^1\,\X\right]_{b\,k}(0)\left[\bar\X\,\gamma^1\,\gamma^3\,n\cdot\A_{hc}\,\right]_{a\,l}(x)|\Omega\rangle\hspace{1em}\Big].
\end{eqnarray}
Comparing (\ref{eq:Jn_repeat}) and (\ref{eq:Tother_pt}) and using $\gamma^3\,\gamma^1\,\gamma^\mu\,\gamma^1\,\gamma^3=\left(\gamma^\mu\right)^T$ we find that
\begin{equation}
T_{\rm other}=\left(\gamma^3\,\gamma^1\,\frac{\nslash}{2}\,\gamma^1\,\gamma^3\,\right)_{ba}\,\delta_{kl}\,{\cal J}_{n}(p^2)=
\left(\frac{\nslash}{2}\right)_{ab}\,\delta_{kl}\,{\cal J}_{n}(p^2),
\end{equation}
which proves equation (\ref{eq:Jn_def}).

As a second example, consider ${\cal J}^S_{11}$ and ${\cal J}^A_{11}$.
Here we use the same transformations to prove the decomposition the
TOP into two jet functions.  In section \ref{subsub:J11} we defined
\begin{eqnarray}
T_{11}&=& \int\,d^4x\, e^{-ip\cdot x}\langle\Omega|\,
T\Big\{\Big[\A^\mu_{\perp
hc}\,\X\Big]_{a\,k}(0)\,,\Big[\bar\X\,\A^\nu_{\perp\,hc}\Big]_{b\,l}(x)\Big\}|\Omega\rangle=
\nonumber\\&& \bar n\cdot
p\,\frac{g_\perp^{\mu\nu}}{d-2}\,\delta_{kl}\,{\cal
J}^S_{11}(p^2)\left(\frac{\nslash}{2}\right)_{ab}+\bar n\cdot
p\,\frac{i\epsilon_\perp^{\mu\nu}}{d-2}\,\delta_{kl}\,{\cal
J}^A_{11}(p^2)\left(\frac{\nslash}{2}\,\gamma_5\right)_{ab}.
\end{eqnarray}
Using translation invariance and the $PT$ symmetry we have
\begin{eqnarray}
T_{11}&=& \int\,d^4x\, e^{-ip\cdot x}\langle\Omega|\,
T\Big\{\Big[\A^\nu_{\perp\,hc}\gamma^3\,\gamma^1\,\X\Big]_{b\,l}(0),\Big[\bar\X\,\gamma^1\,\gamma^3\,\A^\mu_{\perp
hc}\Big]_{a\,k}(0)\Big\}|\Omega\rangle=
\nonumber\\&=& \bar n\cdot
p\,\frac{g_\perp^{\mu\nu}}{d-2}\,\delta_{kl}\,{\cal
J}^S_{11}(p^2)\left(\gamma^3\,\gamma^1\,\frac{\nslash}{2}\,\gamma^1\,\gamma^3\right)_{ba}+\bar n\cdot
p\,\frac{i\epsilon_\perp^{\nu\mu}}{d-2}\,\delta_{kl}\,{\cal
J}^A_{11}(p^2)\left(\gamma^3\,\gamma^1\,\frac{\nslash}{2}\,\gamma_5\,\gamma^1\,\gamma^3\right)_{ba}=\nonumber\\
&=&\bar n\cdot p\,\frac{g_\perp^{\mu\nu}}{d-2}\,\delta_{kl}\,{\cal
J}^S_{11}(p^2)\left(\frac{\nslash}{2}\right)_{ab}+\bar n\cdot
p\,\frac{i\epsilon_\perp^{\mu\nu}}{d-2}\,\delta_{kl}\,{\cal
J}^A_{11}(p^2)\left(\frac{\nslash}{2}\,\gamma_5\right)_{ab},
\end{eqnarray}
which proves that $g_\perp^{\mu\nu}$ is accompanied by $(\nslash/2)$,
while $i\epsilon_\perp^{\mu\nu}$ is accompanied by
$(\nslash\gamma_5/2)$. In a similar manner we can use $PT$ symmetry to
analyze the rest of the TOPs.

\section{\boldmath $T\{J^{(1)},J^{(1)}\}$  subleading jet function(s)}
\label{app:B}
In section \ref{subsub:J11} we argued that the TOP of the first order
SCET current with itself, gives rise to only \emph{one} subleading jet
function. The reason is that the inverse derivative is acting on
\emph{all} the hard-collinear fields. On the other hand, we were
forced to defined two different jet function $j_n$ and $j_{n'}$ since
the inverse derivative was only acting on the hard-collinear gluon and
not on the hard-collinear quark. In this appendix we give a rigorous
proof of these facts.

We consider the TOP of two currents: $J_1(x)$ and $\frac{1}{i\bar
n\cdot\partial}J_2(x)$. Define
\begin{eqnarray}
T_{12}&=&i\int\,d^4x\, e^{-ip\cdot x}\langle\Omega|\,
T\left\{J^\dagger_1(0),\,J_2(x)
\right\}|\Omega\rangle\nonumber\\
T_{21}&=&i\int\,d^4x\, e^{-ip\cdot x}\langle\Omega|\,
T\left\{J^\dagger_2(0),\,J_1(x)
\right\}|\Omega\rangle\nonumber\\
T_{12'}&=&i\int\,d^4x\, e^{-ip\cdot x}\langle\Omega|\,
T\left\{J^\dagger_1(0),\,\frac{1}{i\bar
n\cdot\partial}\,J_2(x)
\right\}|\Omega\rangle\nonumber\\
T_{2'1}&=&i\int\,d^4x\, e^{-ip\cdot x}\langle\Omega|\,
T\left\{J^\dagger_2(0)\,\frac{1}{-i\bar
n\cdot\overleftarrow{\partial}}, \,J_1(x)
\right\}|\Omega\rangle.
\end{eqnarray}
We would like to prove that
\begin{equation}
\frac1\pi\,{\rm Im}\,\bigg(T_{12'}+T_{2'1}\bigg)=-\frac1{\bar n\cdot p}\frac1\pi\,{\rm Im}\,\bigg(T_{12}+T_{12}\bigg).
\end{equation}
Consider $T_{12'}$ first. Inserting a complete set of states
$|r\rangle$ we can write it as
\begin{eqnarray}
T_{12'}=i\int\,d^4x\, e^{-ip\cdot x}&\bigg[&\theta(x^0)\sum_r\,\langle\Omega|\,\frac{1}{i\bar
n\cdot\partial}\,J_2(x)|r\rangle\langle r|J^\dagger_1(0)|\Omega\rangle\nonumber\\
&&\pm\theta(-x^0)\sum_r\,\langle\Omega|J^\dagger_1(0)|r\rangle\langle r|\,\frac{1}{i\bar
n\cdot\partial}\,J_2(x)|\Omega\rangle\bigg].
\end{eqnarray}
Using translation invariance we find
\begin{eqnarray}
T_{12'}=i\int\,d^4x\, e^{-ip\cdot x}&\bigg[&\theta(x^0)\sum_r\,\frac{1}{\bar n\cdot r}\,
\langle\Omega|J_2(0)|r\rangle\langle r|J^\dagger_1(0)|\Omega\rangle e^{-irx}\nonumber\\
&&\pm\theta(-x^0)\sum_r\,\frac{1}{(-\bar n\cdot r)}\,\langle\Omega|J^\dagger_1(0)|r\rangle\langle r|J_2(0)|\Omega\rangle e^{irx}\bigg]
\end{eqnarray}
(we use $r$ to denote both the state and its momentum). Using the identity
$$\theta(x^0)=\frac{i}{2\pi}\int d\omega\, \frac{e^{-i\omega x^0}}{\omega+i\epsilon},
$$
and integrating over $x$ and $\omega$, we obtain
\begin{eqnarray}
T_{12'}&=&i^2\sum_r\,
\langle\Omega|J_2(0)|r\rangle\langle r|J^\dagger_1(0)|\Omega\rangle\,\frac{1}{\bar n\cdot r}\,\frac{(2\pi)^3\delta^3(\vec{p}+\vec{r})}{-p_0-r_0+i\epsilon} \nonumber\\
&&\pm i^2\sum_r\,\langle\Omega|J^\dagger_1(0)|r\rangle\langle r|J_2(0)|\Omega\rangle\,\frac{1}{(-\bar n\cdot r)}\,\frac{(2\pi)^3\delta^3(\vec{p}-\vec{r})}
{p_0-r_0+i\epsilon}.
\end{eqnarray}
Repeating the same procedure for $T_{2'1}$ we find
\begin{eqnarray}
T_{2'1}&=&i^2\sum_r\,
\langle\Omega|J_1(0)|r\rangle\langle r|J^\dagger_2(0)|\Omega\rangle\,\frac{1}{\bar n\cdot r}\,\frac{(2\pi)^3\delta^3(\vec{p}+\vec{r})}{-p_0-r_0+i\epsilon} \nonumber\\
&&\pm i^2\sum_r\,\langle\Omega|J^\dagger_1(0)|r\rangle\langle r|J_2(0)|\Omega\rangle\,\frac{1}{(-\bar n\cdot r)}\,\frac{(2\pi)^3\delta^3(\vec{p}-\vec{r})}
{p_0-r_0+i\epsilon}.
\end{eqnarray}
Since that by definition $\langle a|J_i|b\rangle^*=\langle b|J^\dagger_i|a\rangle$, we find that
\begin{eqnarray}
T_{12'}+T_{2'1}&=&i^2
\sum_r\,2\,{\rm Re}\left[\langle\Omega|J_2(0)|r\rangle\langle r|J^\dagger_1(0)|\Omega\rangle\,\frac{1}{\bar n\cdot r}\right]\,\frac{(2\pi)^3\delta^3(\vec{p}+\vec{r})}{-p_0-r_0+i\epsilon} \nonumber\\
&&\pm i^2\sum_r\,2\,{\rm Re}\left[\langle\Omega|J^\dagger_1(0)|r\rangle\langle r|J_2(0)|\Omega\rangle\,\frac{1}{(-\bar n\cdot r)}\right]\,\frac{(2\pi)^3\delta^3(\vec{p}-\vec{r})}
{p_0-r_0+i\epsilon}.
\end{eqnarray}
Using the identity ${\rm Im}\left[1/(u+i\epsilon)\right]=-\pi\,\delta(u)$, we have
\begin{eqnarray}
\frac1\pi\,{\rm Im}\,\bigg(T_{12'}+T_{2'1}\bigg)&=&\frac{1}{(-\bar n\cdot p)}
\sum_r\,2\,{\rm Re}\left[\langle\Omega|J_2(0)|r\rangle\langle r|J^\dagger_1(0)|\Omega\rangle\right]\,(2\pi)^3\delta^4(p+r) \nonumber\\
&&\pm\frac{1}{(-\bar n\cdot p)} \sum_r\,2\,{\rm Re}\left[\langle\Omega|J^\dagger_1(0)|r\rangle\langle r|J_2(0)|\Omega\rangle\,\right]\,(2\pi)^3\delta^4(p-r)\nonumber\\
&=&-\frac1{\bar n\cdot p}\frac1\pi\,{\rm Im}\,\bigg(T_{12}+T_{12}\bigg).
\end{eqnarray}
We should note that throughout the above derivation we have assumed
that the $\bar n$ component of the total momentum of the state is not
zero and therefore its inverse is defined. This assumption follows
from the fact that by definition the hard-collinear states have a ``large"
$\bar n$ component of momentum. In the same way we can prove that

\begin{eqnarray}
&&\frac 1\pi\,{\rm Im}\,\bigg(i\int\,d^4x\, e^{-ip\cdot x}\langle\Omega|\,
T\left\{J^\dagger_2(0)\,\frac{1}{-i\bar
n\cdot\overleftarrow{\partial}},\,\frac{1}{i\bar
n\cdot\partial}\,J_2(x)
\right\}|\Omega\rangle\bigg)=\nonumber\\
&&=\frac1{(\bar n\cdot p)^2}\frac1\pi\,{\rm Im}\,\bigg(i\int\,d^4x\, e^{-ip\cdot x}\langle\Omega|\,
T\left\{J^\dagger_2(0),\,J_2(x)
\right\}|\Omega\rangle\bigg).
\end{eqnarray}
Using these identities we obtain the results of section \ref{subsub:J11}.

\end{appendix}

\end{document}